\documentclass[notitlepage,english,aps,prl,preprint,tightenlines,floatfix,longbibliography]{revtex4-1}
\usepackage{graphicx}
\usepackage{dcolumn}
\usepackage{bm}
\usepackage{color}
\usepackage[usenames,dvipsnames]{xcolor}
\usepackage[colorlinks=true,urlcolor=blue,citecolor=blue, linkcolor = blue]{hyperref}
\usepackage{amssymb,amsmath}

\usepackage[normalem]{ulem}
\usepackage{upgreek}
\usepackage{lipsum} 
\usepackage[draft]{todonotes}   

\begin{document}

\title{Lasing in dark and bright modes of a finite-sized plasmonic lattice}

\author{T. K. Hakala} 
\thanks{The authors contributed equally to this work.}
\affiliation{COMP Centre of Excellence, Department of Applied Physics, Aalto University School of Science, FI-00076 Aalto, Finland\\
paivi.torma@aalto.fi}

\author{H. T. Rekola} 
\thanks{The authors contributed equally to this work.}
\affiliation{COMP Centre of Excellence, Department of Applied Physics, Aalto University School of Science, FI-00076 Aalto, Finland\\
paivi.torma@aalto.fi}

\author{A. I. V\"akev\"ainen} 
\affiliation{COMP Centre of Excellence, Department of Applied Physics, Aalto University School of Science, FI-00076 Aalto, Finland\\
paivi.torma@aalto.fi}

\author{J.-P. Martikainen} 
\affiliation{COMP Centre of Excellence, Department of Applied Physics, Aalto University School of Science, FI-00076 Aalto, Finland\\
paivi.torma@aalto.fi}

\author{M. Ne\v{c}ada}
\affiliation{COMP Centre of Excellence, Department of Applied Physics, Aalto University School of Science, FI-00076 Aalto, Finland\\
paivi.torma@aalto.fi}

\author{A. J. Moilanen} 
\affiliation{COMP Centre of Excellence, Department of Applied Physics, Aalto University School of Science, FI-00076 Aalto, Finland\\
paivi.torma@aalto.fi}

\author{P. T\"orm\"a}
\affiliation{COMP Centre of Excellence, Department of Applied Physics, Aalto University School of Science, FI-00076 Aalto, Finland\\
paivi.torma@aalto.fi}

\maketitle 

\textbf{Lasing at the nanometre scale promises strong light-matter interactions and ultrafast operation. Plasmonic resonances supported by metallic nanoparticles have extremely small mode volumes and high field enhancements, making them an ideal platform for studying nanoscale lasing. At visible frequencies, however, the applicability of plasmon resonances is limited due to strong ohmic and radiative losses. Intriguingly, plasmonic nanoparticle arrays support non-radiative dark modes that offer longer life-times but are inaccessible to far-field radiation. 
Here, we show lasing both in dark and bright modes of an array of silver nanoparticles combined with optically pumped dye molecules. Linewidths of 0.2 nanometers at visible wavelengths and room temperature are observed. Access to the dark modes is provided by a coherent out-coupling mechanism based on the finite size of the array. The results open a route to utilize all modes of plasmonic lattices, also the high-$Q$ ones, for studies of strong light-matter interactions, condensation and photon fluids.}

\subsection{Introduction}

Plasmonic systems offer small mode volumes, ultrafast dynamics and nanoscale operation, which opens new prospects for light-matter interactions. Strong coupling between plasmonic modes and ensembles of emitters has been demonstrated~\cite{hummer_weak_2013,torma_strong_2015}, and strong coupling even at the level of a few emitters has been reached at room temperature~\cite{chikkaraddy_single-molecule_2016, santhosh_vacuum_2016}. Gain assisted propagation of plasmon polaritons, spacing and lasing have been predicted~\cite{bergman_surface_2003,wuestner_overcoming_2010,dridi_lasing_2015,cuerda_theory_2015,ding_low-threshold_2014}
and studied experimentally~\cite{oulton_plasmon_2009,lu_plasmonic_2012,khajavikhan_thresholdless_2012,lu_all-color_2014,sidiropoulos_ultrafast_2014,zhou_lasing_2013,zheludev_lasing_2008,   noginov_demonstration_2009,lu_plasmonic_2012,suh_plasmonic_2012,van_beijnum_surface_2013,Shalaev2013,schokker_lasing_2014,lu_all-color_2014,Shalaev2014,schokker_statistics_2015,yang_real-time_2015}, 
 and the possibility of photon condensates has been proposed~\cite{martikainen_condensation_2014}. The
fundamental tradeoff between the confinement of optical fields and losses \cite{khurgin_ultimate_2015} render plasmonic systems inherently lossy. 
This motivates a search for hybrid modes where losses can be reduced by a narrow-linewidth component while still preserving the near-field characteristics of the plasmonic component. 

Lattices of metal nanoparticles support modes called surface lattice resonances (SLRs) which are hybrids of localized nanoparticle surface plasmon resonances and diffracted orders (DOs) of the periodic structure~\cite{zou_silver_2004,garcia_de_abajo_textitcolloquium_2007,kravets_extremely_2008,auguie_collective_2008}. The mode energies, losses and optical density of states can be tuned by lattice and particle geometry. The SLRs combine extreme optical confinement with the narrow linewidths inherent in the DO part of the hybrid. Indeed, strong coupling with organic molecules  in such a system has been observed~\cite{vakevainen_plasmonic_2014,shi_spatial_2014,torma_strong_2015}. Infrared lasing at weak coupling regime has been realized in such lattices~\cite{zhou_lasing_2013,yang_real-time_2015}, with linewidths of less than $1.3\, {\rm nm}$~\cite{zhou_lasing_2013}. As the losses of an optical dipole scale as $\omega^4$, noble metals can suffer considerably  higher losses at visible compared to IR. A key question has been whether similar nanoparticle structures could support lasing also at visible frequencies. Importantly, plasmonic lattices can also support so-called dark SLR modes, whose subradiant character results to significantly higher $Q$-values as compared to their radiant (bright) counterparts~\cite{rodriguez_coupling_2011}. Thus plasmonic dark modes are promising candidates for realizing lasing, single-emitter strong coupling, and photon fluids at visible wavelengths. A central challenge is how to excite and couple out these inherently subradiant modes. 

Here, we experimentally demonstrate lasing at the visible wavelengths in both bright and dark modes of the plasmonic lattice.  A new concept to access the dark modes is introduced, which is based on a gradual, coherent build-up of dipole moments in a finite lattice. 

\subsection{Results}
\subsubsection{Dark and bright modes of an infinite lattice}
\begin{figure}
\includegraphics[width=0.95\columnwidth]{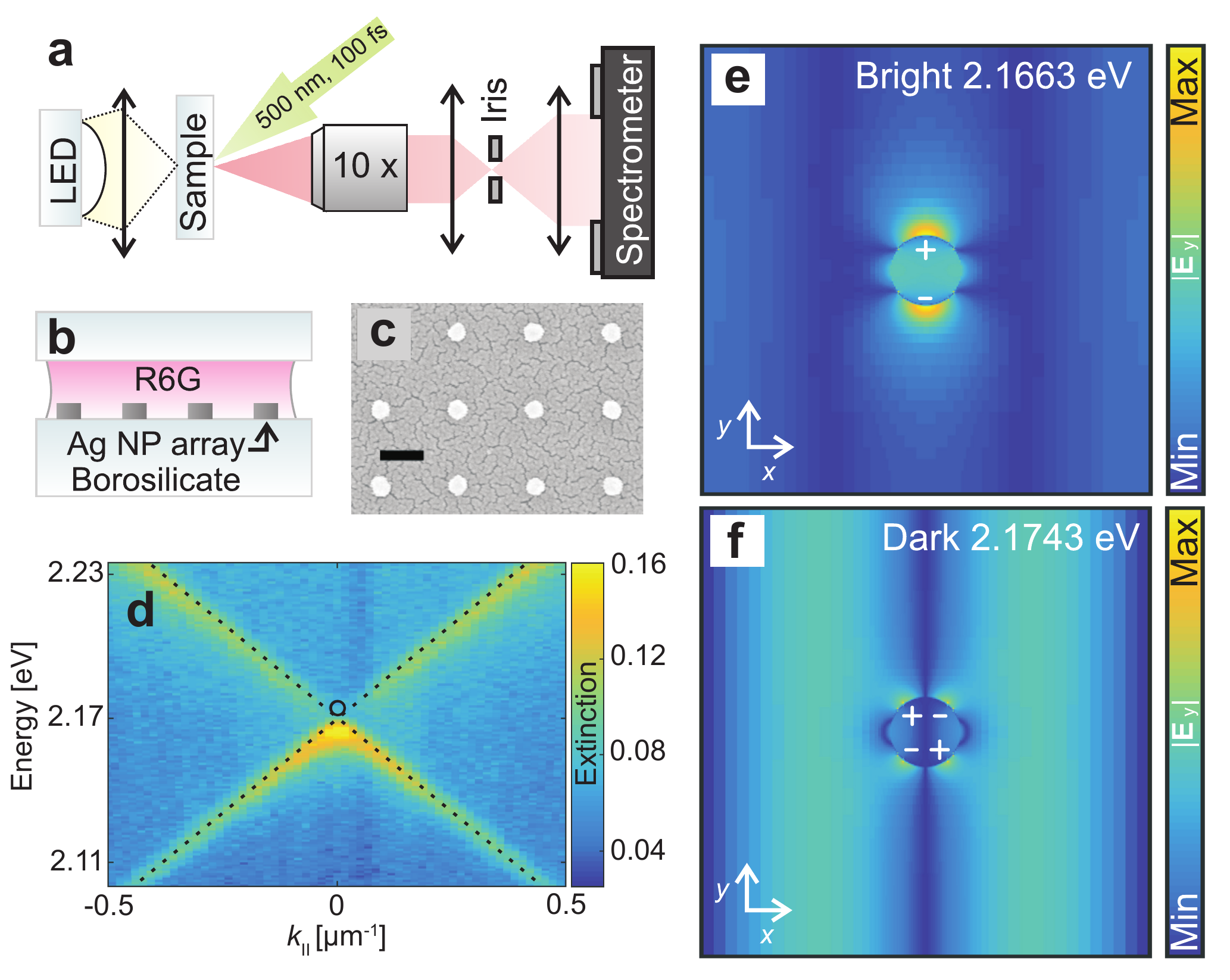}
\caption[Fig1]{{\it The measurement setup, samples, measured dispersion, and field distributions of the SLR modes.} (a) The angle-resolved transmission/luminescence spectra were collected by focusing the image of the backfocal plane of the objective to the entrance slit of the spectrometer. For transmission measurements, a white light source was used. For lasing measurements, the gain medium was pumped with a femtosecond pulsed laser. (b) A schematic of the sample, consisting of rectangular arrays of silver nanoparticles on a borosilicate substrate, a gain medium and a cover glass.  (c) A scanning electron micrograph of a typical sample. The scale bar is $200\, {\rm nm}$. (d) Measured extinction of a typical sample, showing the dispersion of the SLR, which results from  the hybridization of the nanoparticle surface plasmon resonance with the diffracted $<+1, 0>$ and $<-1, 0>$ orders of the lattice (denoted by dashed lines; the crossing of the dashed lines is referred to as the $\Gamma$-point). The location of the dark mode is indicated by the black circle. The colour bar indicates the value of extinction (1-transmission). (e-f) Charge and field distributions of the bright and dark modes for the infinite lattice, obtained from Finite-Difference Time-Domain (FDTD) simulations with periodic boundary conditions. The character of these modes can be understood by considering an array of nanoparticles with period $p_x = p_y$ mainly polarized along $y$ direction and thus radiating predominantly along $-x$ and $+x$ direction. Two counter-propagating radiation fields result to a standing wave which has either a node or an antinode at each particle location, corresponding to the dark or bright mode at $k = 0$, respectively.}
\label{fig1:setup}
\end{figure}

We fabricate and measure arrays of silver nanoparticles on glass, with interparticle spacings of $360-400\, {\rm nm}$ and array sizes of $100 \times 100 \, ({\rm \upmu m})^2$, see Methods. Figs.~\ref{fig1:setup} a-c show the measurement setup, schematic and scanning electron micrograph of a sample. In Fig.~\ref{fig1:setup} d the measured dispersion $E(k_x)$ of a typical sample is shown, where $k_x$ is the in-plane momentum in $x$-direction. Yellow features correspond to the bright modes that respond to the far-field radiation  incident on the sample in the transmission measurement, while the location of the dark mode is indicated by a circle. Figs.~\ref{fig1:setup} e-f show the corresponding electric field and charge distributions found by Finite-Difference Time-Domain (FDTD) simulations for an infinite array. In Fig.~\ref{fig1:setup} e, a standing wave antinode at each particle location induces a large dipole moment and radiation to the far field, thus it is referred to as a bright mode. In Fig.~\ref{fig1:setup} f, however, the field gradient related to a field node induces a quadrupole moment into each particle resulting to zero net dipole moment and negligible far-field radiation, a feature of a dark mode. For both cases, the plasmonic component of the hybrid is evident from the strong near fields in the particle vicinity. A dark mode here specifically refers to one band-edge mode missing in the far field. There are other categories of dark modes in plasmonics systems which present a subradiant mode, especially in metamaterials with symmetry-breaking features ~\cite{Zhang2008, Liu2012}.

\subsubsection{Lasing}

\begin{figure} 
\includegraphics[width=0.95\columnwidth]{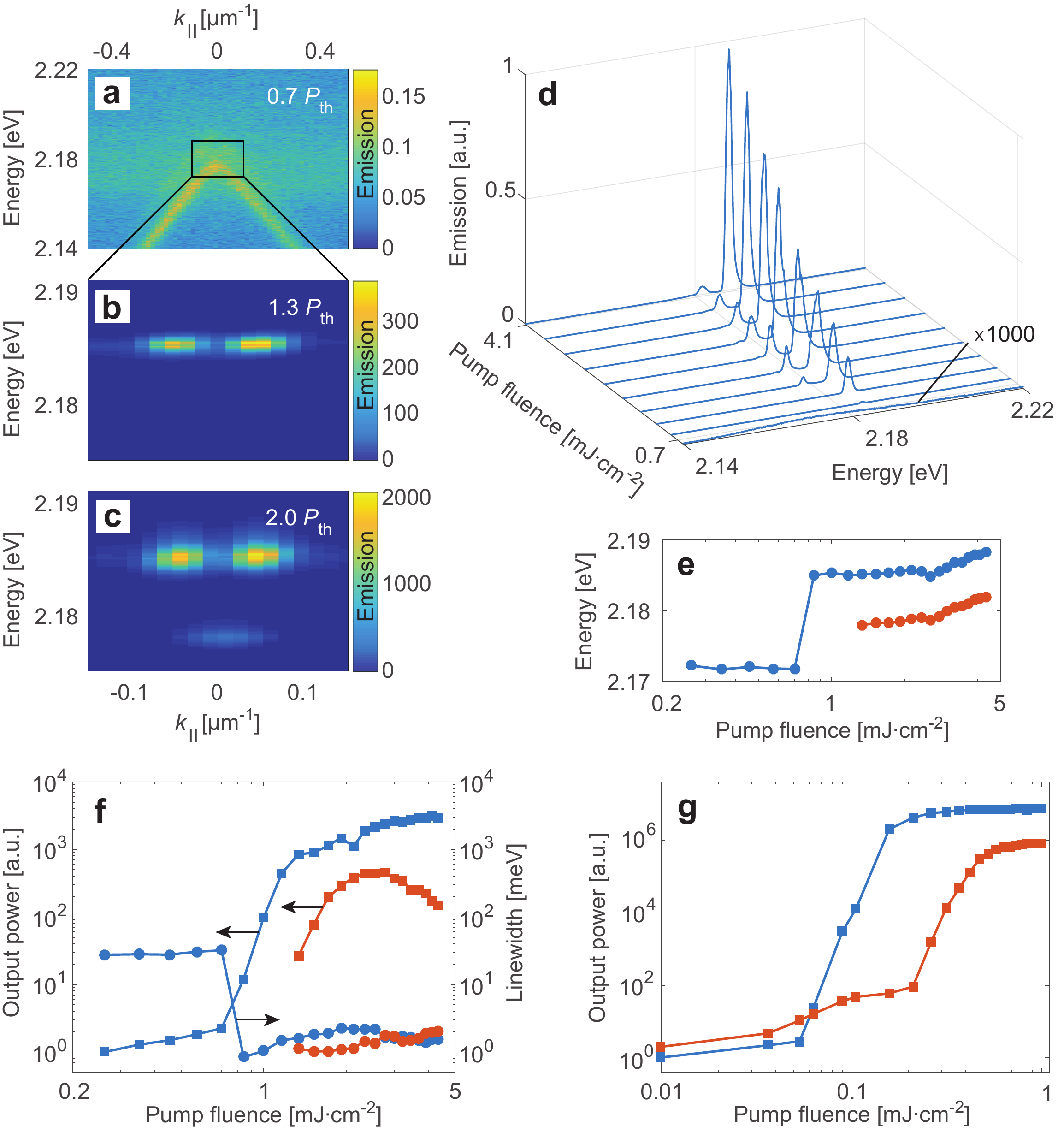}
\caption[Fig2]{{\it Measured emission below and above the lasing threshold, and comparison to numerical results.} Measured emission intensities of a $100 \times 100$ ($\upmu$m)$^2$ nanoparticle array (array periods $p_x = p_y = 375\,{\rm nm}$, particle diameter $d = 60\,{\rm nm}$) with (a) $P = 0.7\, P_{\rm th}$, (b) $P = 1.3\, P_{\rm th}$ and c) $P = 2.0\, P_{\rm th}$, where $P_{\rm th}$ is the threshold pump fluence for the higher energy lasing mode. In a-c, the colour bars indicate the emission intensity in arbitrary units. (d) Emission spectra, (e) the mode energies, and (f) the mode output powers of the lower (red squares) and higher (blue squares) energy modes as a function of pump fluence. Also shown are the linewidths for lower (higher) energy modes as red (blue) dots. (g) FDTD simulation with the gain medium described as four-level systems, the red (blue) squares corresponding to the lower (higher) energy mode, respectively. We have observed lasing for array periodicities $370-390\, {\rm nm}$. Above threshold, we obtain $10^4$ increase in emission intensity, linewidths of 0.2~nm and remarkably low beam divergence of  $\Delta \theta_\mathrm{FWHM} =  0.3^{\circ}$.}
\label{fig2:threshold}
\end{figure}

We combine the lattices with a 31 mM solution of Rhodamine 6G molecules. For concentrations of this order of magnitude, formation of aggregates is not significant~\cite{Hakala_PRL}. Addition of the dye molecules modifies the refractive index surrounding the structure, shifting the SLR resonances slightly from Fig.\ref{fig1:setup}. We excite the molecules with a 100 femtosecond laser pulse (500 nm centre wavelength) and observe the emission, see Methods. Figs.~\ref{fig2:threshold}~a-c show the momentum and energy distribution of the sample emission with different pump fluencies. Below threshold, the emission closely follows the dispersion of the SLR mode (Fig.~\ref{fig2:threshold}~a). The emission spectra in Fig.~\ref{fig2:threshold}~d shows that at threshold, an intense and narrow emission peak at $2.185\, {\rm eV}$ is observed, and well above threshold a second peak appears at $2.178\, {\rm eV}$. The momentum distributions of these two peaks are shown in Figs. ~\ref{fig2:threshold} b and c, respectively. For both peaks, characteristic signatures of lasing, such as rapid nonlinear increase of emission intensity, reduction of linewidth, and a blue shift of the emission are observed in response to increasing pump fluence, see Figs.~\ref{fig2:threshold} e-f. Further, a high beam directionality with divergence of $0.3^\circ$ is observed for both modes. The qualitative behaviour of lasing thresholds is well reproduced by an FDTD simulation with the dye molecules described as four-level systems (for details see Methods), as shown by the Fig.~\ref{fig2:threshold}~g. The simulations show lasing at two separate energies, $2.169\, {\rm eV}$ and $2.142\, {\rm eV}$. The higher energy mode has a lower lasing threshold, analogous to the experiments.


\subsubsection{Control experiments}

To further characterize the lasing, several control experiments were carried out. First, a sample with equal area and amount of nanoparticles but at random positions exhibits no lasing, ruling out random lasing~\cite{wiersma_physics_2008}. FDTD simulations of our structures show that the near fields decay, in the z-direction perpendicular to the array, to $5\%$ of the peak value within $100\, {\rm nm}$. We deposited polymer (PVA) layers of various thicknesses on top of the array so that the gain medium was separated from the nanoparticles: for thickness of about $100\, {\rm nm}$ we observed lasing, but not for $500\, {\rm nm}$ (see Supplementary Fig. 1 and Supplementary Note 1). For phenomena reliant on strong coupling, presence of the emitters extremely close to the nanoparticles is expected to be more essential than for our lasing which takes place in the weak coupling regime. As one more control, in contrast to Ref.~\cite{schokker_statistics_2015} where SLRs hybridize with waveguide modes, our samples exhibit no lasing when over $75 \%$ of particles are removed. For more information about the control experiments with $0 \%$, $50 \%$ and $75 \%$ of particles removed, see Supplementary Figs. 2-4 and Supplementary Note 1.    

\subsubsection{Real space images}

\begin{figure} 
\includegraphics[width=0.8\columnwidth]{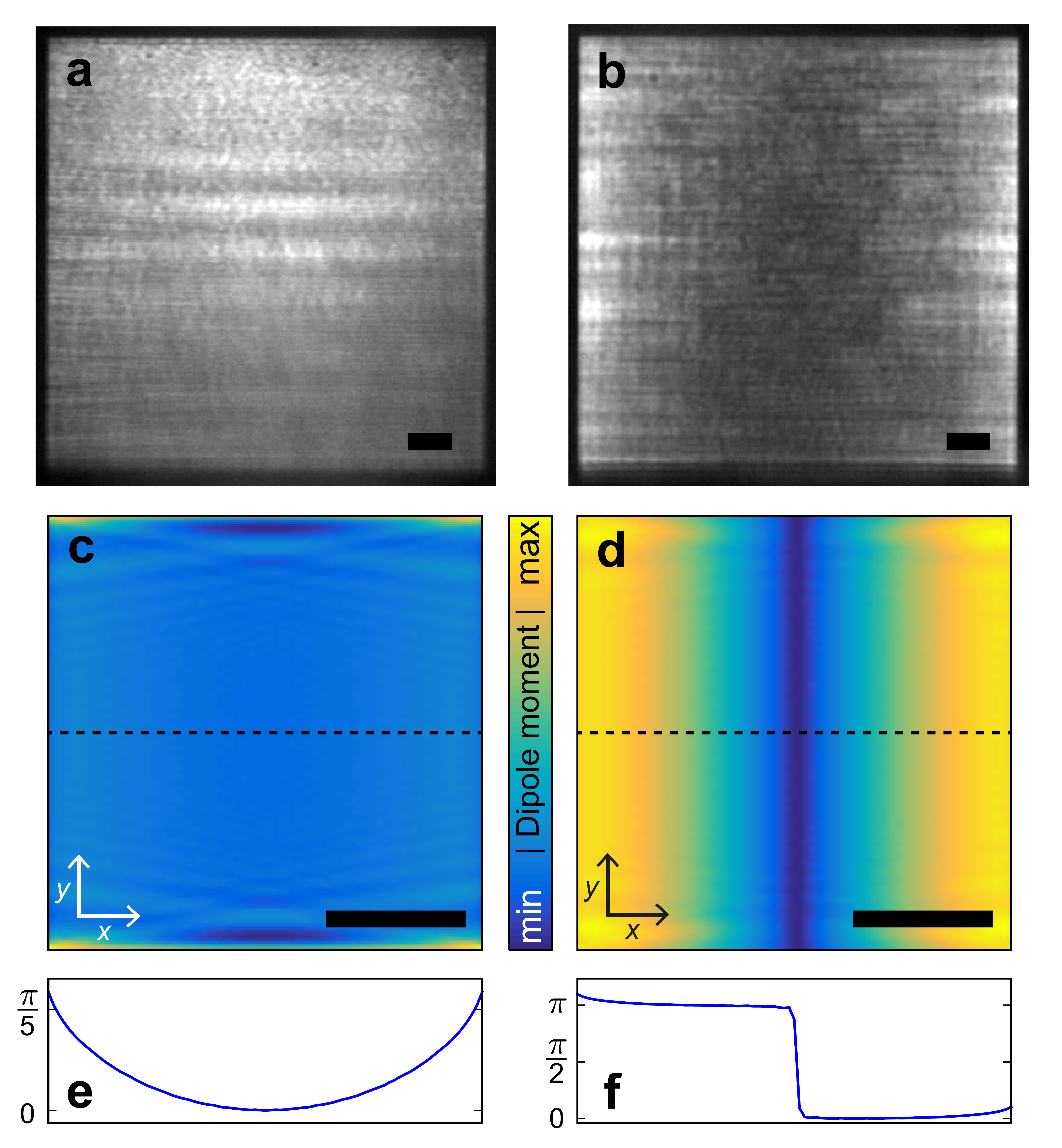}
\caption[Fig4]{{\it Real space images of the lasing action and multipole expansion of scattering for $y$-polarized planewave} ~\cite{xu_radiative_2003}. (a-b) Examples of experimentally observed real space images of the lower (a) and upper (b) energy lasing modes from two different samples. (c-d) Magnitude of the dipole moments (indicated by the colour bar) as a function of position in an array of $85\times 85$ silver nanoparticles, computed using a multipole expansion method without a gain medium, at lower (c) and higher (d) mode energies. The calculated phase of the dipole moments along the dashed lines for both modes are shown in (e) and (f). The scale bars in (a-d) are 10 ${\rm \upmu m}$.} 
\label{fig4}
\end{figure}

From Figs.~\ref{fig2:threshold}~d-f we note that the higher energy lasing mode has a lower threshold, higher emission intensity, and a slightly narrower linewidth compared to the lower energy peak. It could therefore be associated with the low loss dark mode. However, in case of an infinite array size, the dark mode should not radiate to the far field, see Fig.~\ref{fig1:setup} f. Curiously, the higher energy mode exhibits multiple peaks in $k$, whereas the lower energy lasing mode shows only a single peak at $k = 0$, see Fig.~\ref{fig2:threshold}~c. A striking difference between the two cases is observed also in real space images of the lasing action. These far field microscope images correspond to the real space distribution of radiation from the sample but are not the same as the near-field distribution. Figs.~\ref{fig4} a-b show that the lower energy mode lases in the middle of the array, while in the case of the upper energy mode the lasing light is predominantly visible at the edges. 
In Figs.~\ref{fig4} c-d are shown multipolar scattering simulations, without a gain medium, of a finite $85 \times 85$ particle array under $y$-polarized plane wave excitation (for details see Methods). In the simulations the particle polarizabilities include both dipolar and quadrupolar components. The real-space images in far field result from radiation emitted by the sample. The far field emission power is dominated by dipoles, thus the radiated far field intensities in Figs.~\ref{fig4} a-b are proportional to the square of induced dipole moments. Notably, the phase of the dipole moment stays nearly constant across the array for the lower energy mode (Fig.~\ref{fig4}~e), whereas for the higher energy mode the phase undergoes an abrupt $\pi$-phase shift at the centre of the array (Fig.~\ref{fig4}~f). The amplitudes and phases of the quadrupolar moments for both modes are shown in Supplementary Fig. 5 and the related discussion can be found from Supplementary Note 2. 

\subsubsection{Finite size effects}

\begin{figure}
\includegraphics[width=0.6\columnwidth]{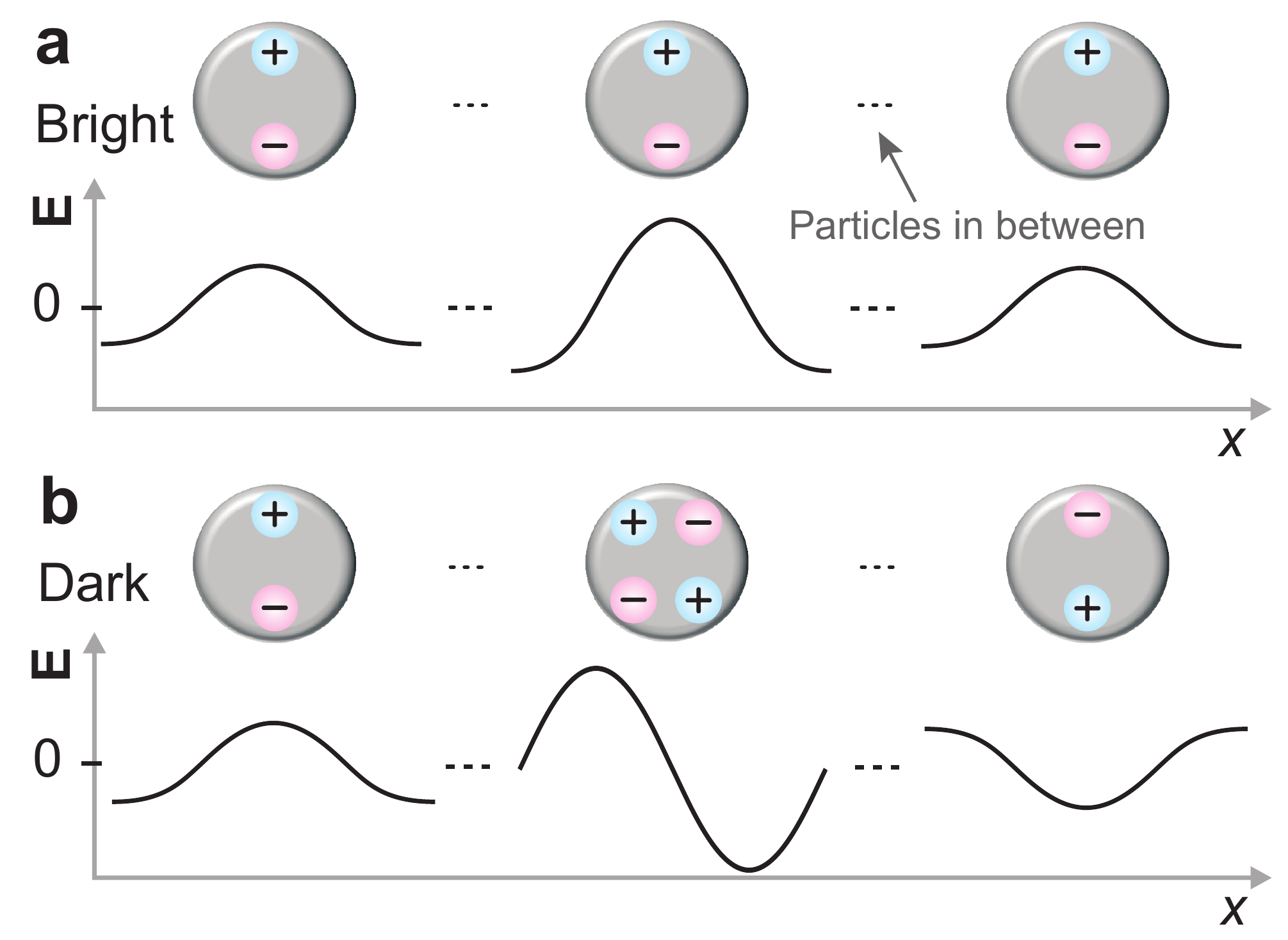}
\caption[Fig4]{{\it Evolution of the radiation fields in the case of a finite array size.} (a) For a bright mode, the phase across the array remains constant due to constructive interference of the counter-propagating radiation fields at each particle location. The amplitude at the edge, however, will be reduced to one half as there are no particles and therefore no radiation incident from the other side. (b) For a dark mode, particles in the array are driven by left- and right-propagating radiation fields, which destructively interfere at the particle locations and therefore create a standing wave node and a quadrupole excitation. As one moves away from the centre, however, the destructive interference becomes gradually less complete due to unequal number of particles contributing to left- and right-propagating waves. This results in a buildup of dipole moments whose relative weight compared to the quadrupoles gradually increases when moving away from the centre of the array. Importantly, for the dark mode the dipoles induced by the left- and right-propagating radiation fields are out of phase by $\pi$. The dark mode in a finite lattice is thus different from the one in an infinite lattice: it features dipole moments and can thus be viewed as a hybrid mode with characteristics of both the bright and dark modes of an infinite lattice.}
\label{fig3}
\end{figure}

We hypothesize that the higher energy mode is indeed the dark mode, and that its visibility, the peculiar beating pattern in $k$ and the real space distribution of the lasing light are all consequences of gradual evolution of radiation fields in the array due to finite-size effects. Fig.~\ref{fig3} shows schematically how finite size effects can lead to radiating dipoles near the edge of the array that are either in the same (bright mode) or opposite (dark mode) phase. Importantly, such a picture holds only if spatial coherence is preserved over the whole array. The profoundly different character of the two modes is manifested in their dipole moment distributions. According to the argument of Fig.~\ref{fig3}, for the bright (dark) mode, the induced dipole moments maximize in the centre (edges) of the array. The dark mode in a finite lattice is thus a hybrid mode containing also features typical for a bright mode of an infinite lattice, such as dipoles, which makes it visible in the far field. The expected radiation in real space is strongest where the dipoles maximize. Indeed, in the real space images of the lasing action, the bright mode mainly emits from the centre of the array (Fig.~\ref{fig4}~a), whereas the dark mode radiates from the edges (Fig.~\ref{fig4}~b). Similar behaviour is visible in the simulations (Figs.~\ref{fig4}~c-d). 

The different real-space distributions of the two lasing modes are expected to lead to distinct behaviour of the corresponding far-field momentum distribution.
Analogous to single slit interference, the bright mode acts as a source with a constant phase and nearly equal amplitude, leading to an odd beating pattern in $k$ (constructive interference at $k=0$).
For the dark mode, radiation maximizes towards the edges, and thus a double slit analogy can be made. Due to the $\pi$-phase shift in the radiation fields across the array, an even beating pattern in $k$ is expected (destructive interference at $k = 0$). Indeed, in Fig.~\ref{fig2:threshold}~c, we observe a peak at $k=0$ for the lower energy mode and two peaks shifted away from $k=0$ for the higher energy, as expected for bright and dark modes, respectively.  

\begin{figure} 
\includegraphics[width=1\columnwidth]{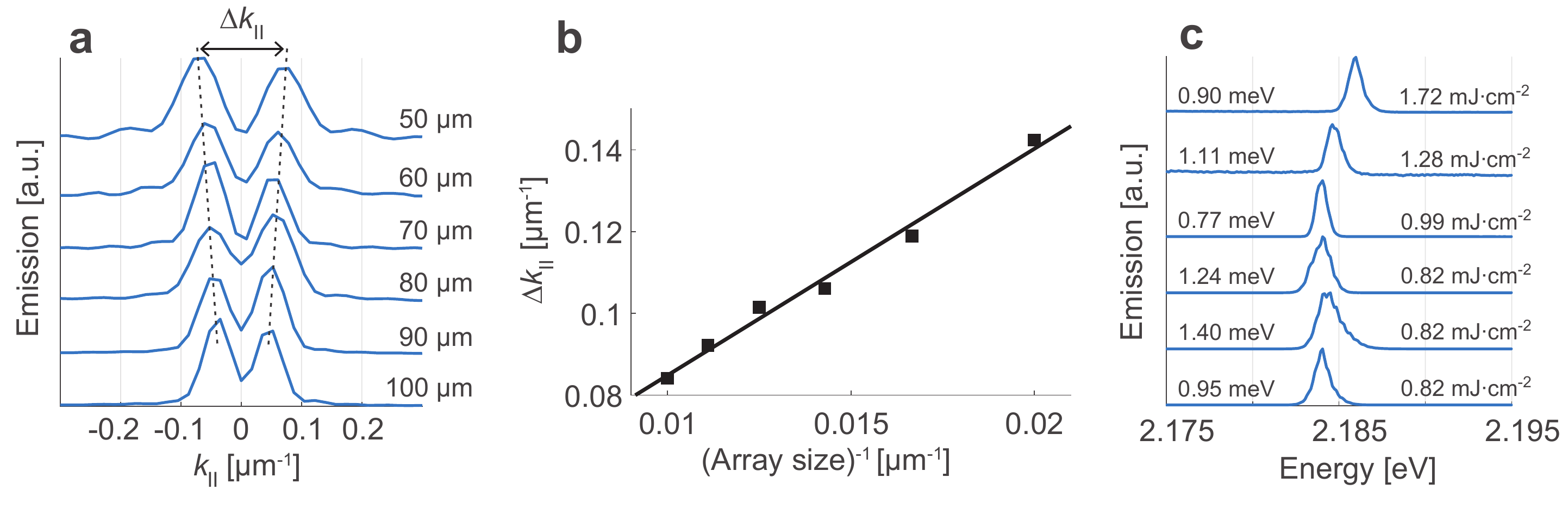}
\caption[Fig5]{{\it Dependence of the k-space beating pattern on the array size.} (a) Normalized constant energy crosscut of the dark mode lasing peak as a function of in-plane $k$-vector for different array sizes. (b) The separation of the peaks in (a) as a function of 1/(array size). The separation of the lasing peaks for 100~${\rm \upmu m}$ and 50~${\rm \upmu m}$ array corresponds to angular separation of 0.44$^{\circ}$ and 0.74$^{\circ}$, respectively. Divergence of individual peaks vary between 0.29$^{\circ}$ and 0.44$^{\circ}$. (c) The lasing spectra for each array size integrated over the double peak from $k = -0.2$ to $k = 0.2\, {\rm \upmu m}^{-1}$. The threshold fluence $P_{\rm th}$ depends on the array size. The linewidth of the lasing peak is denoted on the left and $P_{\rm th}$ on the right.}
\label{fig5}
\end{figure} 

As the ultimate test to our hypothesis, the even beating pattern in $k$ for the dark mode should depend on the array size. This is exactly what we observe: as the array size is decreased, the spacing of the beating pattern maxima gradually increases (Fig.~\ref{fig5}), depending linearly on the inverse of the array size, which is expected for small angles. The associated beating pattern in $k$ shows that spatial coherence of the lasing action extends over the whole array for the system sizes ($50$--$100\, {\rm \upmu m}$) considered. From the lasing spectra for different array sizes (Fig.~\ref{fig5}~c) it is apparent that while smaller arrays have a higher pump threshold, the mode linewidth is not increased significantly. We have thus shown lasing in the dark mode which can be coupled out by a unique approach based on the finite size of the array. 

\section{Discussion}

In summary, we have demonstrated lasing at visible wavelengths both in dark and bright modes of a nanoparticle array. Even though plasmonic systems at visible wavelengths are lossy, we achieve remarkably narrow ($0.2\, {\rm nm}$) linewidths, an increase of four orders of magnitude in emission intensity above threshold, low beam divergence of $0.3^\circ$, and spatial coherence lengths of $100\, {\rm \upmu m}$. This achievement, together with earlier reaching of the strong coupling regime~\cite{vakevainen_plasmonic_2014,shi_spatial_2014}, paves the way for studies of phenomena which combine strong interactions with macroscopic mode population, for example thresholdless lasing, photon and polariton condensation, and photon fluids. The dark-mode out-coupling mechanism that we introduce is not simple scattering or leakage from sharp edges of the system, rather, it is gradual, coherent build-up of dipole moments and radiation intensity. This inspires ideas for the design of not only out-coupling schemes but also beam guiding, trap potentials, topologically non-trivial lattices and edge modes, for instance by gradually changing the pitch and by particle shapes supporting higher order multipoles.

\section{Methods}

\subsubsection{Sample fabrication}

Using electron beam lithography, rectangular arrays of cylindrical (nomimal diameter $60\, {\rm nm}$, height $30\, {\rm nm}$) silver nanoparticles were fabricated on a borosilicate substrate. The angle resolved transmission spectra were obtained by focusing light from a white LED onto a sample. A $10\times$ 0.3 NA objective focused onto the sample surface collected the transmitted light.  

\subsubsection{Measurement setup}

The image of the back focal plane of the objective was focused to the entrance slit of the spectrometer. Thus the transmitted (or emitted) light from the sample into a given angle is focused into a single position on the entrance slit. One axis of the 2-dimensional CCD camera of the spectrometer can then be used to resolve the angle of the transmitted /emitted light, while the other axis is used to resolve the wavelength. The $E(k)$ dispersions were subsequently calculated from the angle and wavelength resolved spectra as $E=hc/\lambda$ and $k_{||}=k_0 \operatorname{sin}(\theta)$. Here $k_0=2\pi/\lambda$ and $\theta$ is the angle with respect to sample normal. The real space images were obtained by focusing the image plane of the same sample onto a separate 2-D CCD camera.

For lasing measurements, the gain medium was pumped with a femtosecond laser ($45^\circ$ incident angle, $500\, {\rm nm}$ centre wavelength, $100\, {\rm fs}$ pulse width, $1\, {\rm kHz}$ repetition rate). The gain medium consisted of $31\, {\rm mM}$ Rhodamine 6G in Benzyl Alcohol.

\subsubsection{Data analysis}

The beam divergence is obtained from the angle resolved lasing data (see Fig.~\ref{fig2:threshold}) by plotting the emission intensity as a function of in-plane k-vector at the lasing energy. From this crosscut, we can determine the full-width-half-maximum value of each individual lasing peak, $\Delta k_\mathrm{FWHM}$. For small angles $\operatorname{sin}(\theta) \approx \theta$, and thus the divergence can finally be defined as $\Delta \theta_\mathrm{FWHM} = \Delta k_\mathrm{FWHM} / k_0$ [rad], where $k_0$ is the free space wavenumber. The angular separation of the two peaks at dark mode energy (see Fig.~\ref{fig5}) can be calculated with the previous equation by replacing $\Delta k_\mathrm{FWHM}$ with peak separation $\Delta k$.

\subsubsection{Simulations}

FDTD simulations were done using Lumerical's FDTD Solutions software. The nanoparticle was modelled as a $30\, {\rm nm}$ tall cylinder with a diameter of $60\, {\rm nm}$. Tabulated material parameters for silver were used \cite{Palik1991}. The background refractive index was set to 1.52. For the field profile simulations the simulation mesh was set to $0.6\, {\rm nm}$ over the volume of the nanoparticle. Ten dipole sources with random orientation, location, and phase excite the SLR modes with a $3.3\, {\rm fs}$ pulse at the beginning of the simulation. Field profiles were recorded after the simulation had ran for $700\, {\rm fs}$ to filter out the excitation pulse. An infinite array was considered in the simulations by choosing the $x$ and $y$ boundary conditions to be either asymmetric and symmetric (bright mode, Fig.~\ref{fig1:setup} e), or asymmetric and asymmetric (dark mode, Fig.~\ref{fig1:setup} f), respectively. This selection of boundary conditions allows to isolate the modes of interest, and control the polarization direction of the bright mode.  As the structure has the same periodicity in $x$ and $y$, equivalent modes exist also in the perpendicular direction to the one shown in Figs.~\ref{fig1:setup} e and f.

The simulations with gain in FDTD were done using a similar model, except this time periodic boundary conditions were used for both $x$ and $y$ to allow both the bright and dark modes to exist simultaneously. The excitation of the modes is done by pumping the dye molecules with a $500\,{\rm nm}$, $30\,{\rm fs}$, $0.01$--$2.5\,{\rm mJ\cdot cm^{-2}}$ pulse normally incident on the structure. The dye is modelled as four-level systems, which are coupled to the electric fields in time domain, similar to Ref.~\cite{DridiSchatz2013paper}. After the pump pulse, the excitations in the dye molecules relax to the upper lasing state, and from there excite the bright and dark SLR modes through stimulated emission if the population inversion is sufficient to overcome the losses in the system, see Supplementary Figs. 6-7 and Supplementary Note 3. For the parameters used to model the dye and for a more detailed description of the model, see see Supplementary Fig. 8 and Supplementary Note 3. In addition to the FDTD modelling shown in the main text, we have used a neo-classical model with simplified description for the SLR modes to simulate the lasing action in a nanoparticle array \cite{martikainen_modelling_2016}, see Supplementary Fig. 9 and Supplementary note 4 for details. 

\subsubsection{Theoretical calculations}

Our theoretical calculations of the array response, see Fig.~\ref{fig4}, are based on the multiple scattering T-matrix method~\cite{mishchenko_t-matrix_1999}, which is a generalization of the often-used coupled dipole approximation, enabling to take into account multipole responses of a single nanoparticle up to an arbitrary multipolar order. We compute the electromagnetic response ($T$-matrix) of a single nanoparticle using {\sc scuff-em}, an open-source implementation of the boundary-element method (http://github.com/homerreid/scuff-EM)~\cite{SCUFF1}. The nanoparticle parameters are the same as in the FDTD simulations above. All the calculations are performed up to the quadrupole order. Fig.~\ref{fig4}~c shows the magnitudes of the electric dipole moment in the $y$-direction among the nanoparticles when the array is excited by the $y$-polarized plane wave coming from the normal direction, at a frequency corresponding to the bright mode. Fig.~\ref{fig4}~d depicts the same $y$-dipole response when the array is excited by two $y$-polarized plane waves with opposite incident angle, corresponding to the peaks of the observed $k$-space beating patterns. A brief explanation of the multiple scattering method and figures showing the quadrupole response among the nanoparticles are provided in the Supplementary Fig. 5 and Supplementary Note 2.

\subsubsection{Data availability}
The authors declare that relevant data supporting the findings of this study are available on request.

\section{acknowledgments}
This work was supported by the Academy of Finland through its  Centres  of  Excellence  Programme (2012--2017)  and under  Project  Nos.  263347,  284621, and  272490,  and by  the  European  Research  Council  (ERC-2013-AdG-340748-CODE). This article is based upon work from COST Action MP1403 “Nanoscale Quantum Optics,” supported by COST (European Cooperation in Science and Technology). Part of the research was performed at the Micronova Nanofabrication Centre, supported by Aalto University. Triton cluster  at Aalto University was used for the computations. The authors thank M. Dridi for fruitful discussions.

\section{Author contributions}
P.T. initiated and supervised the project. T.K.H., H.T.R. and A.I.V. designed and performed the experiments and analysed the data. H.T.R. and J.-P. M. performed the FDTD simulations. M. N., J.-P.M. and A.J.M. did the numerical modelling of the finite-sized arrays. H.T.R., T.K.H. and A.I.V. fabricated the samples. All authors discussed the results. T.K.H. and P.T. wrote the manuscript together with all authors.

\section{Competing financial interests:}
The authors declare no competing financial interests.


\begin{thebibliography}{10}
\expandafter\ifx\csname url\endcsname\relax
  \def\url#1{\texttt{#1}}\fi
\expandafter\ifx\csname urlprefix\endcsname\relax\def\urlprefix{URL }\fi
\providecommand{\bibinfo}[2]{#2}
\providecommand{\eprint}[2][]{\url{#2}}

\bibitem{hummer_weak_2013}
\bibinfo{author}{H{\"u}mmer, T.}, \bibinfo{author}{Garc{\'i}a-Vidal, F.~J.},
  \bibinfo{author}{Mart{\'i}n-Moreno, L.} \& \bibinfo{author}{Zueco, D.}
\newblock \bibinfo{title}{Weak and strong coupling regimes in plasmonic {QED}}.
\newblock \emph{\bibinfo{journal}{Phys. Rev. B}} \textbf{\bibinfo{volume}{87}},
  \bibinfo{pages}{115419} (\bibinfo{year}{2013}).

\bibitem{torma_strong_2015}
\bibinfo{author}{T\"{o}rm\"{a}, P.} \& \bibinfo{author}{Barnes, W.~L.}
\newblock \bibinfo{title}{Strong coupling between surface plasmon polaritons
  and emitters: a review}.
\newblock \emph{\bibinfo{journal}{Rep. Prog. Phys.}}
  \textbf{\bibinfo{volume}{78}}, \bibinfo{pages}{013901}
  (\bibinfo{year}{2015}).

\bibitem{chikkaraddy_single-molecule_2016}
\bibinfo{author}{Chikkaraddy, R.} \emph{et~al.}
\newblock \bibinfo{title}{Single-molecule strong coupling at room temperature
  in plasmonic nanocavities}.
\newblock \emph{\bibinfo{journal}{Nature}} \textbf{\bibinfo{volume}{535}},
  \bibinfo{pages}{127--130} (\bibinfo{year}{2016}).

\bibitem{santhosh_vacuum_2016}
\bibinfo{author}{Santhosh, K.}, \bibinfo{author}{Bitton, O.},
  \bibinfo{author}{Chuntonov, L.} \& \bibinfo{author}{Haran, G.}
\newblock \bibinfo{title}{Vacuum {Rabi} splitting in a plasmonic cavity at the
  single quantum emitter limit}.
\newblock \emph{\bibinfo{journal}{Nat. Commun.}} \textbf{\bibinfo{volume}{7}},
  \bibinfo{pages}{11823} (\bibinfo{year}{2016}).

\bibitem{bergman_surface_2003}
\bibinfo{author}{Bergman, D.~J.} \& \bibinfo{author}{Stockman, M.~I.}
\newblock \bibinfo{title}{Surface {Plasmon} {Amplification} by {Stimulated}
  {Emission} of {Radiation}: {Quantum} {Generation} of {Coherent} {Surface}
  {Plasmons} in {Nanosystems}}.
\newblock \emph{\bibinfo{journal}{Phys. Rev. Lett.}}
  \textbf{\bibinfo{volume}{90}}, \bibinfo{pages}{027402}
  (\bibinfo{year}{2003}).

\bibitem{wuestner_overcoming_2010}
\bibinfo{author}{Wuestner, S.}, \bibinfo{author}{Pusch, A.},
  \bibinfo{author}{Tsakmakidis, K.~L.}, \bibinfo{author}{Hamm, J.~M.} \&
  \bibinfo{author}{Hess, O.}
\newblock \bibinfo{title}{Overcoming {Losses} with {Gain} in a {Negative}
  {Refractive} {Index} {Metamaterial}}.
\newblock \emph{\bibinfo{journal}{Phys. Rev. Lett.}}
  \textbf{\bibinfo{volume}{105}}, \bibinfo{pages}{127401}
  (\bibinfo{year}{2010}).

\bibitem{dridi_lasing_2015}
\bibinfo{author}{Dridi, M.} \& \bibinfo{author}{Schatz, G.~C.}
\newblock \bibinfo{title}{Lasing action in periodic arrays of nanoparticles}.
\newblock \emph{\bibinfo{journal}{J. Opt. Soc. Am. B}}
  \textbf{\bibinfo{volume}{32}}, \bibinfo{pages}{818} (\bibinfo{year}{2015}).

\bibitem{cuerda_theory_2015}
\bibinfo{author}{Cuerda, J.}, \bibinfo{author}{R{\"{u}}ting, F.},
  \bibinfo{author}{Garc\'{i}a-Vidal, F.~J.} \& \bibinfo{author}{Bravo-Abad, J.}
\newblock \bibinfo{title}{Theory of lasing action in plasmonic crystals}.
\newblock \emph{\bibinfo{journal}{Phys. Rev. B}} \textbf{\bibinfo{volume}{91}},
  \bibinfo{pages}{041118} (\bibinfo{year}{2015}).

\bibitem{ding_low-threshold_2014}
\bibinfo{author}{Ding, P.} \emph{et~al.}
\newblock \bibinfo{title}{Low-threshold resonance amplification of out-of-plane
  lattice plasmons in active plasmonic nanoparticle arrays}.
\newblock \emph{\bibinfo{journal}{J. Opt.}} \textbf{\bibinfo{volume}{16}},
  \bibinfo{pages}{065003} (\bibinfo{year}{2014}).

\bibitem{oulton_plasmon_2009}
\bibinfo{author}{Oulton, R.~F.} \emph{et~al.}
\newblock \bibinfo{title}{Plasmon lasers at deep subwavelength scale}.
\newblock \emph{\bibinfo{journal}{Nature}} \textbf{\bibinfo{volume}{461}},
  \bibinfo{pages}{629--632} (\bibinfo{year}{2009}).

\bibitem{lu_plasmonic_2012}
\bibinfo{author}{Lu, Y.-J.} \emph{et~al.}
\newblock \bibinfo{title}{Plasmonic {Nanolaser} {Using} {Epitaxially} {Grown}
  {Silver} {Film}}.
\newblock \emph{\bibinfo{journal}{Science}} \textbf{\bibinfo{volume}{337}},
  \bibinfo{pages}{450--453} (\bibinfo{year}{2012}).

\bibitem{khajavikhan_thresholdless_2012}
\bibinfo{author}{Khajavikhan, M.} \emph{et~al.}
\newblock \bibinfo{title}{Thresholdless nanoscale coaxial lasers}.
\newblock \emph{\bibinfo{journal}{Nature}} \textbf{\bibinfo{volume}{482}},
  \bibinfo{pages}{204--207} (\bibinfo{year}{2012}).

\bibitem{lu_all-color_2014}
\bibinfo{author}{Lu, Y.-J.} \emph{et~al.}
\newblock \bibinfo{title}{All-{Color} {Plasmonic} {Nanolasers} with {Ultralow}
  {Thresholds}: {Autotuning} {Mechanism} for {Single}-{Mode} {Lasing}}.
\newblock \emph{\bibinfo{journal}{Nano Lett.}} \textbf{\bibinfo{volume}{14}},
  \bibinfo{pages}{4381--4388} (\bibinfo{year}{2014}).

\bibitem{sidiropoulos_ultrafast_2014}
\bibinfo{author}{Sidiropoulos, T. P.~H.} \emph{et~al.}
\newblock \bibinfo{title}{Ultrafast plasmonic nanowire lasers near the surface
  plasmon frequency}.
\newblock \emph{\bibinfo{journal}{Nat. Phys.}} \textbf{\bibinfo{volume}{10}},
  \bibinfo{pages}{870--876} (\bibinfo{year}{2014}).

\bibitem{zhou_lasing_2013}
\bibinfo{author}{Zhou, W.} \emph{et~al.}
\newblock \bibinfo{title}{Lasing action in strongly coupled plasmonic
  nanocavity arrays}.
\newblock \emph{\bibinfo{journal}{Nat. Nanotechnol.}}
  \textbf{\bibinfo{volume}{8}}, \bibinfo{pages}{506--511}
  (\bibinfo{year}{2013}).

\bibitem{zheludev_lasing_2008}
\bibinfo{author}{Zheludev, N.~I.}, \bibinfo{author}{Prosvirnin, S.~L.},
  \bibinfo{author}{Papasimakis, N.} \& \bibinfo{author}{Fedotov, V.~A.}
\newblock \bibinfo{title}{Lasing spaser}.
\newblock \emph{\bibinfo{journal}{Nat. Photonics}}
  \textbf{\bibinfo{volume}{2}}, \bibinfo{pages}{351--354}
  (\bibinfo{year}{2008}).

\bibitem{noginov_demonstration_2009}
\bibinfo{author}{Noginov, M.~A.} \emph{et~al.}
\newblock \bibinfo{title}{Demonstration of a spaser-based nanolaser}.
\newblock \emph{\bibinfo{journal}{Nature}} \textbf{\bibinfo{volume}{460}},
  \bibinfo{pages}{1110--1112} (\bibinfo{year}{2009}).

\bibitem{suh_plasmonic_2012}
\bibinfo{author}{Suh, J.~Y.} \emph{et~al.}
\newblock \bibinfo{title}{Plasmonic {Bowtie} {Nanolaser} {Arrays}}.
\newblock \emph{\bibinfo{journal}{Nano Lett.}} \textbf{\bibinfo{volume}{12}},
  \bibinfo{pages}{5769--5774} (\bibinfo{year}{2012}).

\bibitem{van_beijnum_surface_2013}
\bibinfo{author}{van Beijnum, F.} \emph{et~al.}
\newblock \bibinfo{title}{Surface {Plasmon} {Lasing} {Observed} in {Metal}
  {Hole} {Arrays}}.
\newblock \emph{\bibinfo{journal}{Phys. Rev. Lett.}}
  \textbf{\bibinfo{volume}{110}}, \bibinfo{pages}{206802}
  (\bibinfo{year}{2013}).

\bibitem{Shalaev2013}
\bibinfo{author}{Meng, X.}, \bibinfo{author}{Kildishev, A.~V.},
  \bibinfo{author}{Fujita, K.}, \bibinfo{author}{Tanaka, K.} \&
  \bibinfo{author}{Shalaev, V.~M.}
\newblock \bibinfo{title}{Wavelength-tunable spasing in the visible}.
\newblock \emph{\bibinfo{journal}{Nano Lett.}} \textbf{\bibinfo{volume}{13}},
  \bibinfo{pages}{4106--4112} (\bibinfo{year}{2013}).

\bibitem{schokker_lasing_2014}
\bibinfo{author}{Schokker, A.~H.} \& \bibinfo{author}{Koenderink, A.~F.}
\newblock \bibinfo{title}{Lasing at the band edges of plasmonic lattices}.
\newblock \emph{\bibinfo{journal}{Phys. Rev. B}} \textbf{\bibinfo{volume}{90}},
  \bibinfo{pages}{155452} (\bibinfo{year}{2014}).

\bibitem{Shalaev2014}
\bibinfo{author}{Meng, X.}, \bibinfo{author}{Liu, J.},
  \bibinfo{author}{Kildishev, A.~V.} \& \bibinfo{author}{Shalaev, V.~M.}
\newblock \bibinfo{title}{Highly directional spaser array for the red
  wavelength region}.
\newblock \emph{\bibinfo{journal}{Laser Photonics Reviews}}
  \textbf{\bibinfo{volume}{8}}, \bibinfo{pages}{896--903}
  (\bibinfo{year}{2014}).

\bibitem{schokker_statistics_2015}
\bibinfo{author}{Schokker, A.~H.} \& \bibinfo{author}{Koenderink, A.~F.}
\newblock \bibinfo{title}{Statistics of {Randomized} {Plasmonic} {Lattice}
  {Lasers}}.
\newblock \emph{\bibinfo{journal}{ACS Photonics}} \textbf{\bibinfo{volume}{2}},
  \bibinfo{pages}{1289--1297} (\bibinfo{year}{2015}).

\bibitem{yang_real-time_2015}
\bibinfo{author}{Yang, A.} \emph{et~al.}
\newblock \bibinfo{title}{Real-time tunable lasing from plasmonic nanocavity
  arrays}.
\newblock \emph{\bibinfo{journal}{Nat. Commun.}} \textbf{\bibinfo{volume}{6}},
  \bibinfo{pages}{6939} (\bibinfo{year}{2015}).

\bibitem{martikainen_condensation_2014}
\bibinfo{author}{Martikainen, J.-P.}, \bibinfo{author}{Heikkinen, M. O.~J.} \&
  \bibinfo{author}{T\"{o}rm\"{a}, P.}
\newblock \bibinfo{title}{Condensation phenomena in plasmonics}.
\newblock \emph{\bibinfo{journal}{Phys. Rev. A}} \textbf{\bibinfo{volume}{90}},
  \bibinfo{pages}{053604} (\bibinfo{year}{2014}).

\bibitem{khurgin_ultimate_2015}
\bibinfo{author}{Khurgin, J.~B.}
\newblock \bibinfo{title}{Ultimate limit of field confinement by surface
  plasmon polaritons}.
\newblock \emph{\bibinfo{journal}{Faraday Discuss.}}
  \textbf{\bibinfo{volume}{178}}, \bibinfo{pages}{109--122}
  (\bibinfo{year}{2015}).

\bibitem{zou_silver_2004}
\bibinfo{author}{Zou, S.}, \bibinfo{author}{Janel, N.} \&
  \bibinfo{author}{Schatz, G.~C.}
\newblock \bibinfo{title}{Silver nanoparticle array structures that produce
  remarkably narrow plasmon lineshapes}.
\newblock \emph{\bibinfo{journal}{J. Chem. Phys.}}
  \textbf{\bibinfo{volume}{120}}, \bibinfo{pages}{10871--10875}
  (\bibinfo{year}{2004}).

\bibitem{garcia_de_abajo_textitcolloquium_2007}
\bibinfo{author}{Garc\'{i}a~de Abajo, F.~J.}
\newblock \bibinfo{title}{{Colloquium}: {Light} scattering by particle and hole
  arrays}.
\newblock \emph{\bibinfo{journal}{Rev. Mod. Phys.}}
  \textbf{\bibinfo{volume}{79}}, \bibinfo{pages}{1267--1290}
  (\bibinfo{year}{2007}).

\bibitem{kravets_extremely_2008}
\bibinfo{author}{Kravets, V.~G.}, \bibinfo{author}{Schedin, F.} \&
  \bibinfo{author}{Grigorenko, A.~N.}
\newblock \bibinfo{title}{Extremely {Narrow} {Plasmon} {Resonances} {Based} on
  {Diffraction} {Coupling} of {Localized} {Plasmons} in {Arrays} of {Metallic}
  {Nanoparticles}}.
\newblock \emph{\bibinfo{journal}{Phys. Rev. Lett.}}
  \textbf{\bibinfo{volume}{101}}, \bibinfo{pages}{087403}
  (\bibinfo{year}{2008}).

\bibitem{auguie_collective_2008}
\bibinfo{author}{Augui\'e, B.} \& \bibinfo{author}{Barnes, W.~L.}
\newblock \bibinfo{title}{Collective {Resonances} in {Gold} {Nanoparticle}
  {Arrays}}.
\newblock \emph{\bibinfo{journal}{Phys. Rev. Lett.}}
  \textbf{\bibinfo{volume}{101}}, \bibinfo{pages}{143902}
  (\bibinfo{year}{2008}).

\bibitem{vakevainen_plasmonic_2014}
\bibinfo{author}{V{\"{a}}kev{\"{a}}inen, A.~I.} \emph{et~al.}
\newblock \bibinfo{title}{Plasmonic {Surface} {Lattice} {Resonances} at the
  {Strong} {Coupling} {Regime}}.
\newblock \emph{\bibinfo{journal}{Nano Lett.}} \textbf{\bibinfo{volume}{14}},
  \bibinfo{pages}{1721--1727} (\bibinfo{year}{2014}).

\bibitem{shi_spatial_2014}
\bibinfo{author}{Shi, L.} \emph{et~al.}
\newblock \bibinfo{title}{Spatial {Coherence} {Properties} of {Organic}
  {Molecules} {Coupled} to {Plasmonic} {Surface} {Lattice} {Resonances} in the
  {Weak} and {Strong} {Coupling} {Regimes}}.
\newblock \emph{\bibinfo{journal}{Phys. Rev. Lett.}}
  \textbf{\bibinfo{volume}{112}}, \bibinfo{pages}{153002}
  (\bibinfo{year}{2014}).

\bibitem{rodriguez_coupling_2011}
\bibinfo{author}{Rodriguez, S. R.~K.} \emph{et~al.}
\newblock \bibinfo{title}{Coupling {Bright} and {Dark} {Plasmonic} {Lattice}
  {Resonances}}.
\newblock \emph{\bibinfo{journal}{Phys. Rev. X}} \textbf{\bibinfo{volume}{1}},
  \bibinfo{pages}{021019} (\bibinfo{year}{2011}).

\bibitem{Zhang2008}
\bibinfo{author}{Zhang, S.}, \bibinfo{author}{Genov, D.~A.},
  \bibinfo{author}{Wang, Y.}, \bibinfo{author}{Liu, M.} \&
  \bibinfo{author}{Zhang, X.}
\newblock \bibinfo{title}{Plasmon-induced transparency in metamaterials}.
\newblock \emph{\bibinfo{journal}{Phys. Rev. Lett.}}
  \textbf{\bibinfo{volume}{101}}, \bibinfo{pages}{047401}
  (\bibinfo{year}{2008}).

\bibitem{Liu2012}
\bibinfo{author}{Liu, X.} \emph{et~al.}
\newblock \bibinfo{title}{Electromagnetically induced transparency in terahertz
  plasmonic metamaterials via dual excitation pathways of the dark mode}.
\newblock \emph{\bibinfo{journal}{Appl. Phys. Lett.}}
  \textbf{\bibinfo{volume}{100}} (\bibinfo{year}{2012}).

\bibitem{Hakala_PRL}
\bibinfo{author}{Hakala, T.~K.} \emph{et~al.}
\newblock \bibinfo{title}{Vacuum {Rabi} {Splitting} and {Strong-Coupling}
  {Dynamics} for {Surface-Plasmon} {Polaritons} and {Rhodamine 6G}
  {Molecules}}.
\newblock \emph{\bibinfo{journal}{Phys. Rev. Lett.}}
  \textbf{\bibinfo{volume}{103}}, \bibinfo{pages}{053602}
  (\bibinfo{year}{2009}).

\bibitem{wiersma_physics_2008}
\bibinfo{author}{Wiersma, D.~S.}
\newblock \bibinfo{title}{The physics and applications of random lasers}.
\newblock \emph{\bibinfo{journal}{Nat. Phys.}} \textbf{\bibinfo{volume}{4}},
  \bibinfo{pages}{359--367} (\bibinfo{year}{2008}).

\bibitem{Palik1991}
\bibinfo{editor}{Palik, E.~D.} (ed.) \emph{\bibinfo{title}{Handbook of Optical
  Constants of Solids}}, vol. \bibinfo{volume}{1--3}
  (\bibinfo{publisher}{Academic Press Inc}, \bibinfo{year}{1991}).

\bibitem{DridiSchatz2013paper}
\bibinfo{author}{Dridi, M.} \& \bibinfo{author}{Schatz, G.~C.}
\newblock \bibinfo{title}{Model for describing plasmon-enhanced lasers that
  combines rate equations with finite-difference time-domain}.
\newblock \emph{\bibinfo{journal}{J. Opt. Soc. Am. B}}
  \textbf{\bibinfo{volume}{30}}, \bibinfo{pages}{2791} (\bibinfo{year}{2013}).

\bibitem{martikainen_modelling_2016}
\bibinfo{author}{Martikainen, J.-P.}, \bibinfo{author}{Hakala, T.~K.},
  \bibinfo{author}{Rekola, H.~T.} \& \bibinfo{author}{T{\"{o}}rm{\"{a}}, P.}
\newblock \bibinfo{title}{Modelling lasing in plasmonic nanoparticle arrays}.
\newblock \emph{\bibinfo{journal}{J. Opt.}} \textbf{\bibinfo{volume}{18}},
  \bibinfo{pages}{024006} (\bibinfo{year}{2016}).

\bibitem{mishchenko_t-matrix_1999}
\bibinfo{author}{Mishchenko, M.~I.}, \bibinfo{author}{Travis, L.~D.} \&
  \bibinfo{author}{Macke, A.}
\newblock \bibinfo{title}{T-matrix method and its applications}.
\newblock In \emph{\bibinfo{booktitle}{Light {Scattering} by {Nonspherical}
  {Particles}: {Theory}, {Measurements}, and {Applications}}},
  \bibinfo{pages}{147--172} (\bibinfo{publisher}{Academic Press},
  \bibinfo{year}{1999}).

\bibitem{SCUFF1}
\bibinfo{author}{{Homer Reid}, M.~T.} \& \bibinfo{author}{{Johnson}, S.~G.}
\newblock \bibinfo{title}{{Efficient Computation of Power, Force, and Torque in
  BEM Scattering Calculations}}.
\newblock \emph{\bibinfo{journal}{IEEE Trans. Antennas Propagat.}}
  \textbf{\bibinfo{volume}{63}}, \bibinfo{pages}{3588--3598}
  (\bibinfo{year}{2015}).

\bibitem{xu_radiative_2003}
\bibinfo{author}{Xu, Y.-l.}
\newblock \bibinfo{title}{Radiative scattering properties of an ensemble of
  variously shaped small particles}.
\newblock \emph{\bibinfo{journal}{Physical Review E}}
  \textbf{\bibinfo{volume}{67}}, \bibinfo{pages}{046620}
  (\bibinfo{year}{2003}).

\end{thebibliography}

\clearpage

\begin{thebibliography}{10}
\expandafter\ifx\csname url\endcsname\relax
  \def\url#1{\texttt{#1}}\fi
\expandafter\ifx\csname urlprefix\endcsname\relax\def\urlprefix{URL }\fi
\providecommand{\bibinfo}[2]{#2}
\providecommand{\eprint}[2][]{\url{#2}}

\bibitem{schokker_statistics_2015}
\bibinfo{author}{Schokker, A.~H.} \& \bibinfo{author}{Koenderink, A.~F.}
\newblock \bibinfo{title}{Statistics of {Randomized} {Plasmonic} {Lattice}
  {Lasers}}.
\newblock \emph{\bibinfo{journal}{ACS Photonics}} \textbf{\bibinfo{volume}{2}},
  \bibinfo{pages}{1289--1297} (\bibinfo{year}{2015}).

\bibitem{mishchenko_t-matrix_1999}
\bibinfo{author}{Mishchenko, M.~I.}, \bibinfo{author}{Travis, L.~D.} \&
  \bibinfo{author}{Macke, A.}
\newblock \bibinfo{title}{T-matrix method and its applications}.
\newblock In \emph{\bibinfo{booktitle}{Light {Scattering} by {Nonspherical}
  {Particles}: {Theory}, {Measurements}, and {Applications}}},
  \bibinfo{pages}{147--172} (\bibinfo{publisher}{Academic Press},
  \bibinfo{year}{1999}).

\bibitem{xu_calculation_1996}
\bibinfo{author}{Xu, Y.-l.}
\newblock \bibinfo{title}{Calculation of the {Addition} {Coefficients} in
  {Electromagnetic} {Multisphere}-{Scattering} {Theory}}.
\newblock \emph{\bibinfo{journal}{J. Comput. Phys.}}
  \textbf{\bibinfo{volume}{127}}, \bibinfo{pages}{285--298}
  (\bibinfo{year}{1996}).

\bibitem{xu_radiative_2003}
\bibinfo{author}{Xu, Y.-l.}
\newblock \bibinfo{title}{Radiative scattering properties of an ensemble of
  variously shaped small particles}.
\newblock \emph{\bibinfo{journal}{Physical Review E}}
  \textbf{\bibinfo{volume}{67}}, \bibinfo{pages}{046620}
  (\bibinfo{year}{2003}).

\bibitem{taylor_optical_2011}
\bibinfo{author}{Taylor, J.~M.}
\newblock \emph{\bibinfo{title}{Optical {Binding} {Phenomena}: {Observations}
  and {Mechanisms}}} (\bibinfo{publisher}{Springer Science \& Business Media},
  \bibinfo{year}{2011}).

\bibitem{SCUFF1}
\bibinfo{author}{{Homer Reid}, M.~T.} \& \bibinfo{author}{{Johnson}, S.~G.}
\newblock \bibinfo{title}{{Efficient Computation of Power, Force, and Torque in
  BEM Scattering Calculations}}.
\newblock \emph{\bibinfo{journal}{IEEE Trans. Antennas Propagat.}}
  \textbf{\bibinfo{volume}{63}}, \bibinfo{pages}{3588--3598}
  (\bibinfo{year}{2015}).

\bibitem{SCUFF2}
 \bibinfo{note}{{http://github.com/homerreid/scuff-EM}}.

\bibitem{DridiSchatz2013paper}
\bibinfo{author}{Dridi, M.} \& \bibinfo{author}{Schatz, G.~C.}
\newblock \bibinfo{title}{Model for describing plasmon-enhanced lasers that
  combines rate equations with finite-difference time-domain}.
\newblock \emph{\bibinfo{journal}{J. Opt. Soc. Am. B}}
  \textbf{\bibinfo{volume}{30}}, \bibinfo{pages}{2791} (\bibinfo{year}{2013}).

\bibitem{dridi_lasing_2015}
\bibinfo{author}{Dridi, M.} \& \bibinfo{author}{Schatz, G.~C.}
\newblock \bibinfo{title}{Lasing action in periodic arrays of nanoparticles}.
\newblock \emph{\bibinfo{journal}{J. Opt. Soc. Am. B}}
  \textbf{\bibinfo{volume}{32}}, \bibinfo{pages}{818} (\bibinfo{year}{2015}).

\bibitem{martikainen_modelling_2016}
\bibinfo{author}{Martikainen, J.-P.}, \bibinfo{author}{Hakala, T.~K.},
  \bibinfo{author}{Rekola, H.~T.} \& \bibinfo{author}{T{\"{o}}rm{\"{a}}, P.}
\newblock \bibinfo{title}{Modelling lasing in plasmonic nanoparticle arrays}.
\newblock \emph{\bibinfo{journal}{J. Opt.}} \textbf{\bibinfo{volume}{18}},
  \bibinfo{pages}{024006} (\bibinfo{year}{2016}).

\bibitem{cuerda_theory_2015}
\bibinfo{author}{Cuerda, J.}, \bibinfo{author}{R{\"{u}}ting, F.},
  \bibinfo{author}{Garc\'{i}a-Vidal, F.~J.} \& \bibinfo{author}{Bravo-Abad, J.}
\newblock \bibinfo{title}{Theory of lasing action in plasmonic crystals}.
\newblock \emph{\bibinfo{journal}{Phys. Rev. B}} \textbf{\bibinfo{volume}{91}},
  \bibinfo{pages}{041118} (\bibinfo{year}{2015}).

\end{thebibliography}

\clearpage
\setcounter{figure}{0} 
\def\bibsection{\subsection*{Supplementary References.}} 

\section{Supplementary information}

\maketitle

\renewcommand{\figurename}{\textbf{Supplementary Figure}}

\begin{figure}[h!]
\includegraphics[width=1\columnwidth]{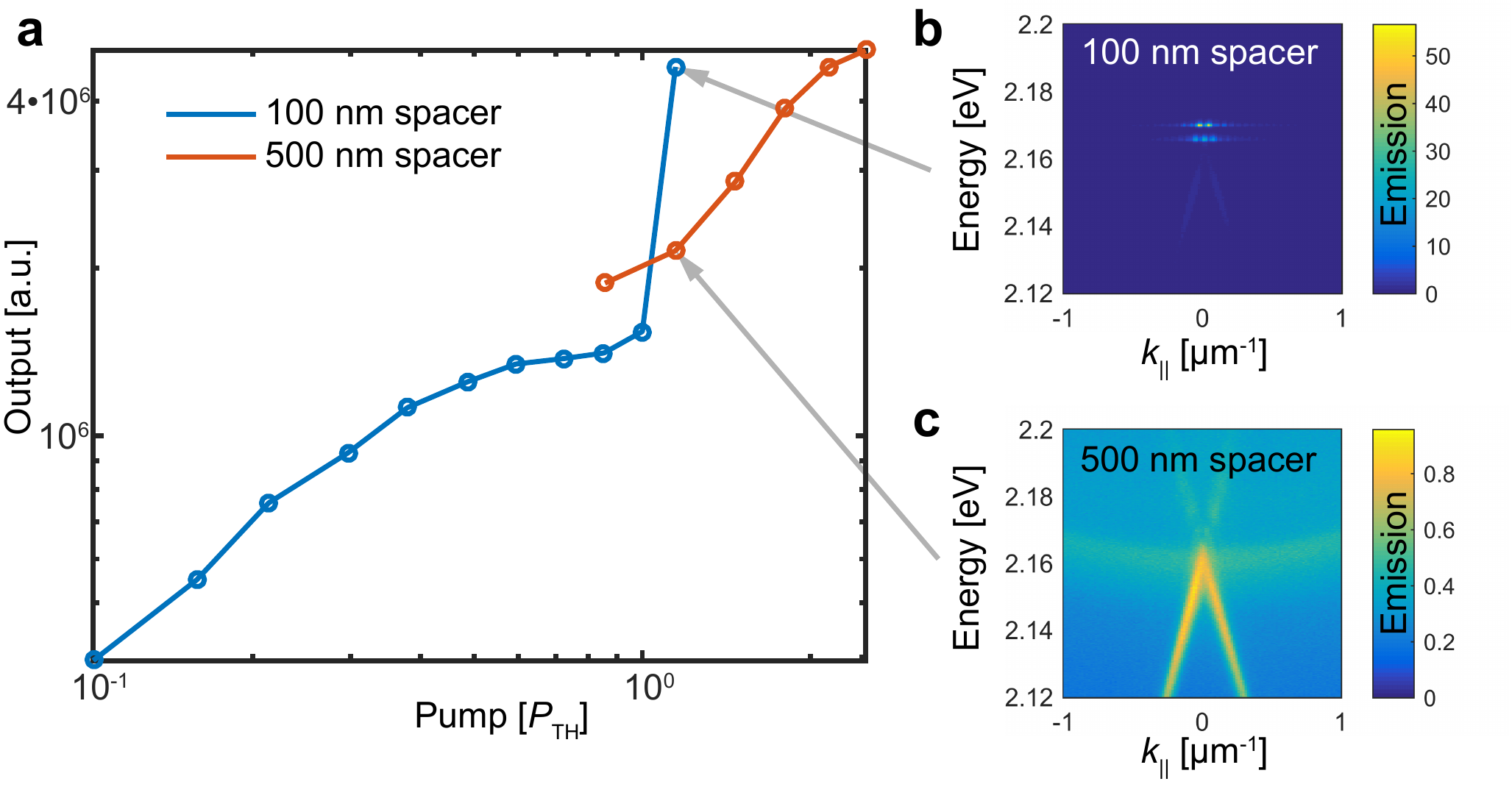}
\caption{{\it Control experiment results where a dielectric spacer layer was added on top of the nanoparticle array.} Luminescence intensity as a function of the pump power (a) for samples having a PVA spacer layer of 100 nm (blue) and 500 nm (red). The pump powers are normalized with the threshold of the sample having the 100 nm PVA layer. Luminescence as a function of in-plane wave vector and energy at $P_{\mathrm{TH}}$ for a PVA spacer of 100 nm (b) and 500 nm (c). The colorscale represents the observed emission intensity in (b) and (c) at the slightly above the threshold pump power $P_{\mathrm{TH}}$. \label{fig:PVA_spacer_thresholds}}
\end{figure}

\begin{figure}[hb!]
\includegraphics[width=\columnwidth]{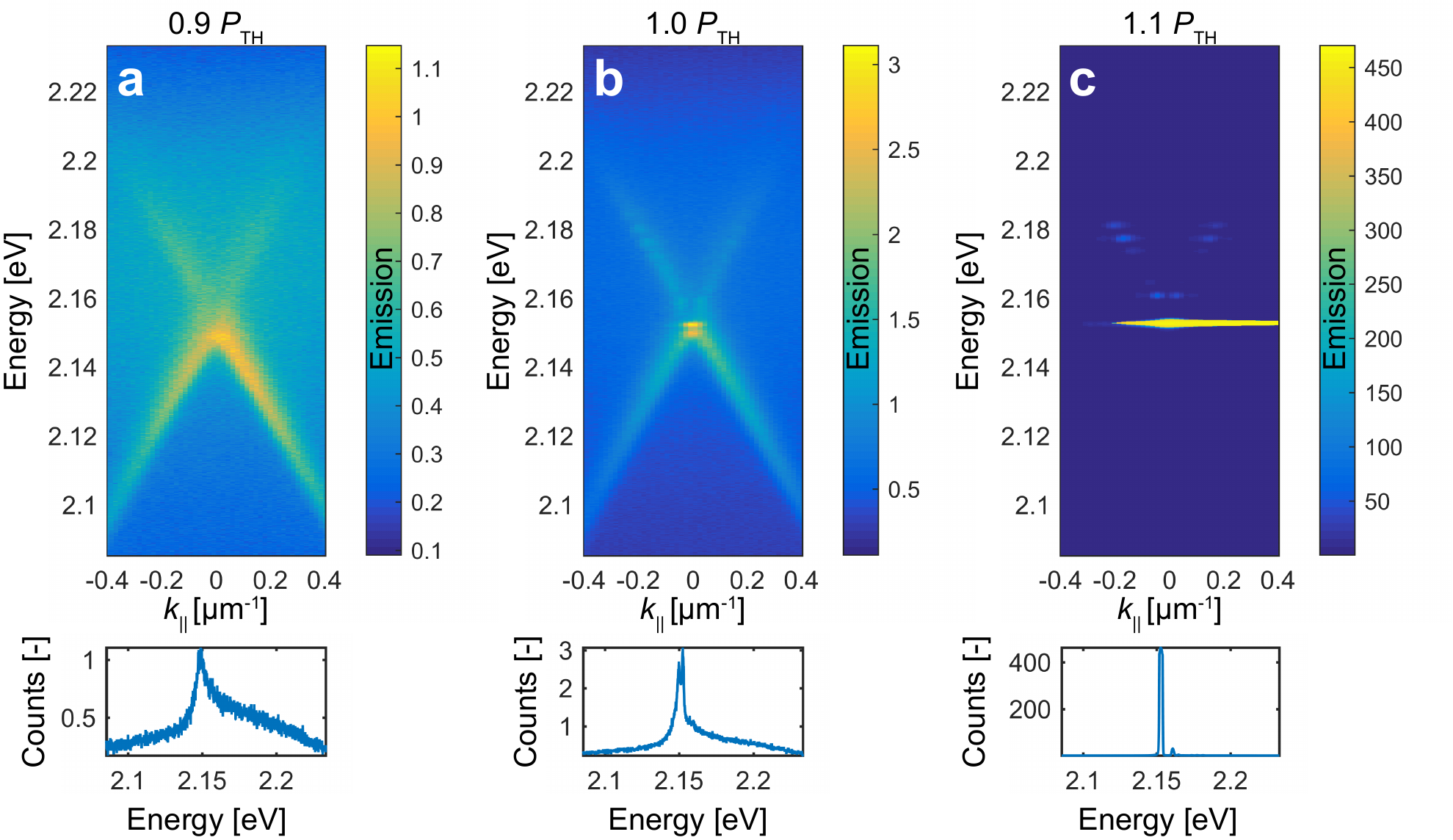}
\caption{{\it Control experiment results where a fraction of the nanoparticles are removed.} (a-c) Luminescence data for a sample with 0 \% of particles removed and with pump powers 0.9 $P_{\mathrm{TH}}$, 1.0 $P_{\mathrm{TH}}$ and 1.1 $P_{\mathrm{TH}}$, respectively. The colorscale represents the observed emission intensity. In (c), the spectrometer CCD was strongly saturated by the laser emission, causing the apparent spreading along the horizontal axis. Below each plot are shown the crosscuts in energy at $k_{||} = 0$. \label{fig:removed particles0}}
\end{figure}

\begin{figure}[h!]
\includegraphics[width=\columnwidth]{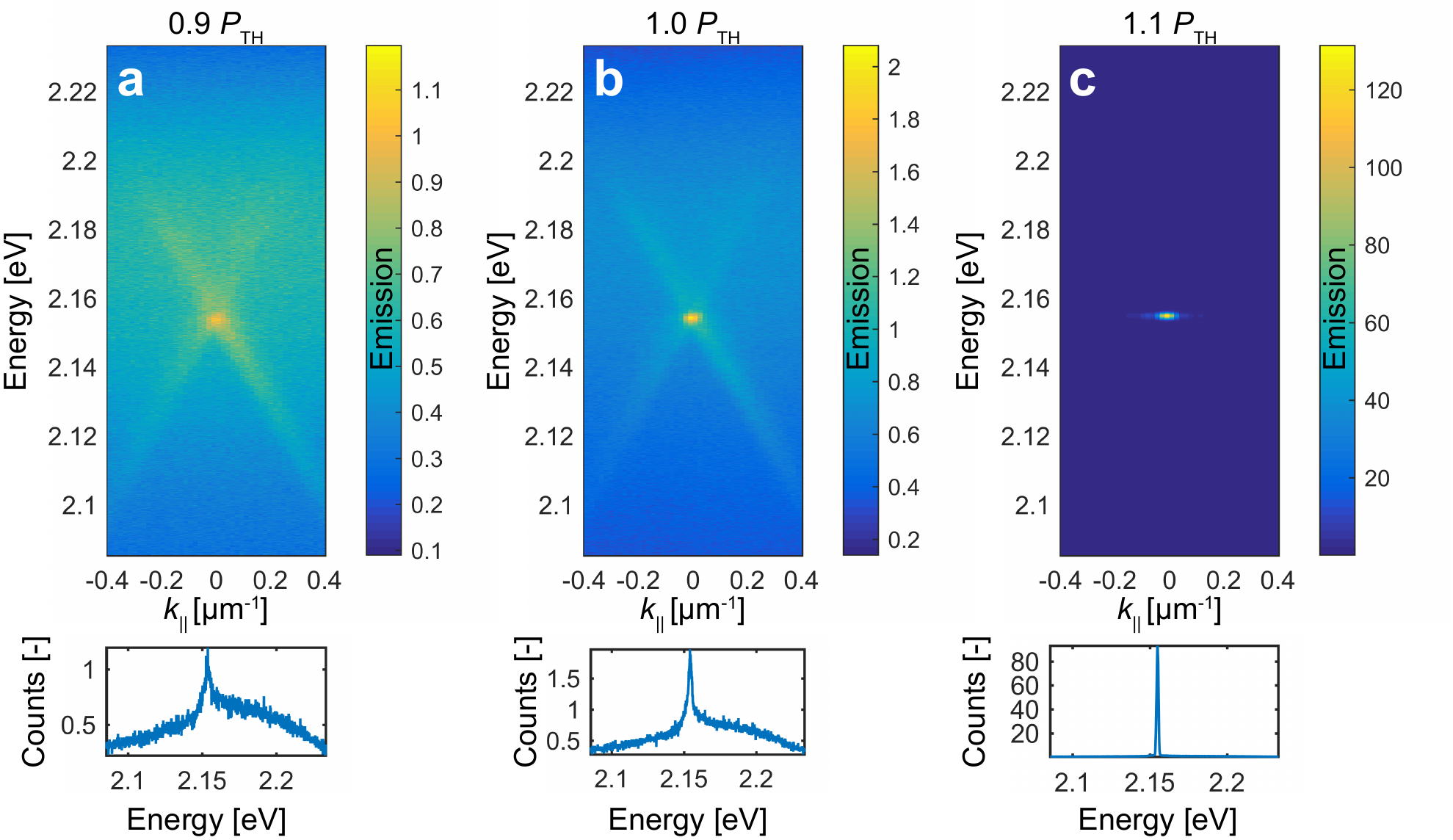}
\caption{{\it Control experiment results where a fraction of the nanoparticles are removed.} Luminescence data for a sample with 50 \% of particles removed from random positions on the array. The pump powers were (a) 0.9 $P_{\mathrm{TH}}$, (b) 1.0 $P_{\mathrm{TH}}$ and (c) 1.1 $P_{\mathrm{TH}}$. The colorscale represents the observed emission intensity. Below each plot are shown the crosscuts in energy at $k_{||} = 0$.
\label{fig:removed particles50}}
\end{figure}

\begin{figure}[h!]
\includegraphics[width=\columnwidth]{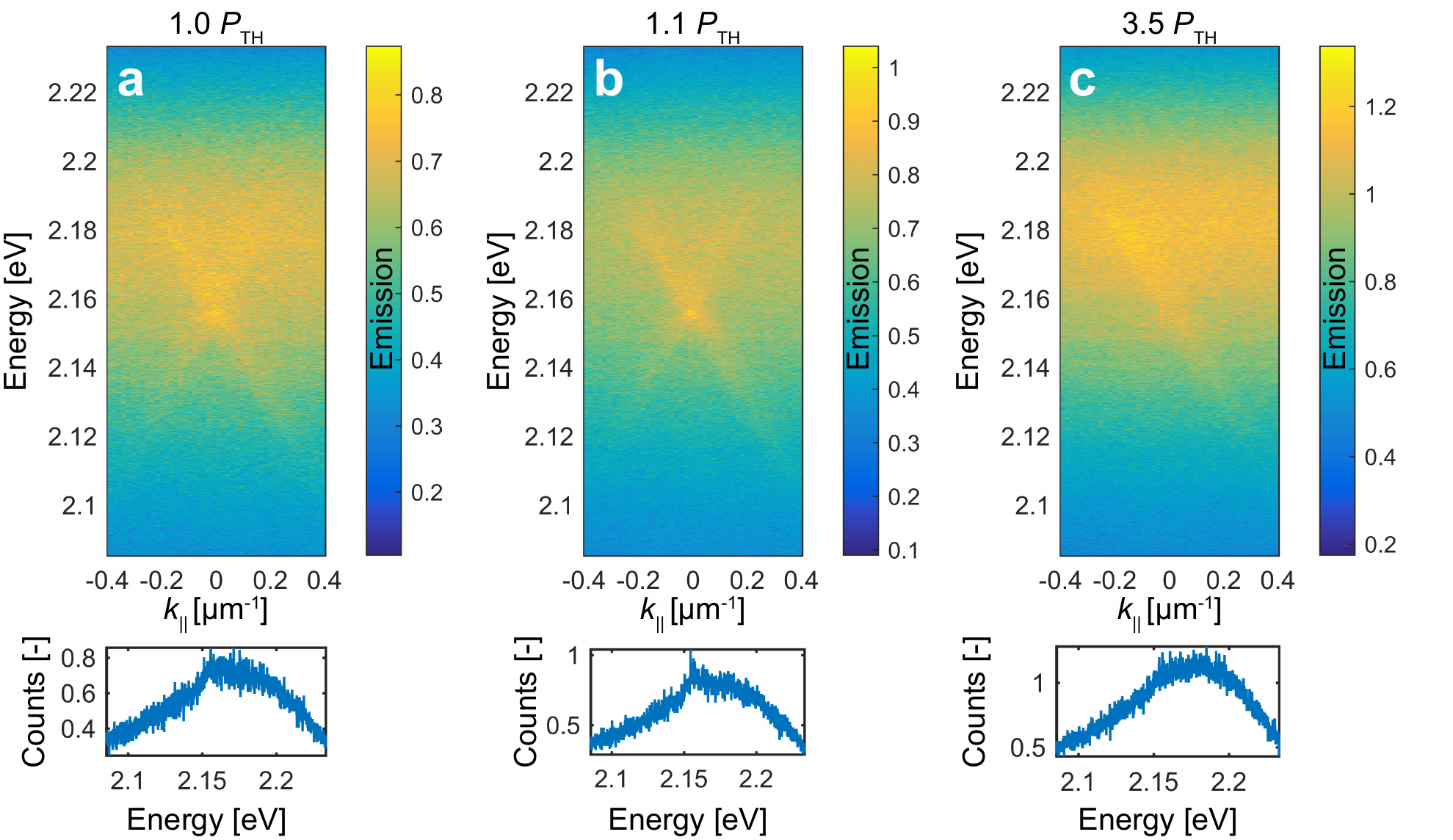}
\caption{{\it Control experiment results where a fraction of the nanoparticles are removed.} Luminescence data for a sample with 75 \% of particles removed from random positions on the array. The pump powers were (a) 1.0 $P_{\mathrm{TH}}$, (b) 1.1 $P_{\mathrm{TH}}$ and (c) 3.5 $P_{\mathrm{TH}}$. The colorscale represents the observed emission intensity. Below each plot are shown the crosscuts in energy at $k_{||} = 0$. No lasing was observed. \label{fig:removed particles75}}
\end{figure}


\begin{figure}[h!]
\includegraphics[width=1\columnwidth]{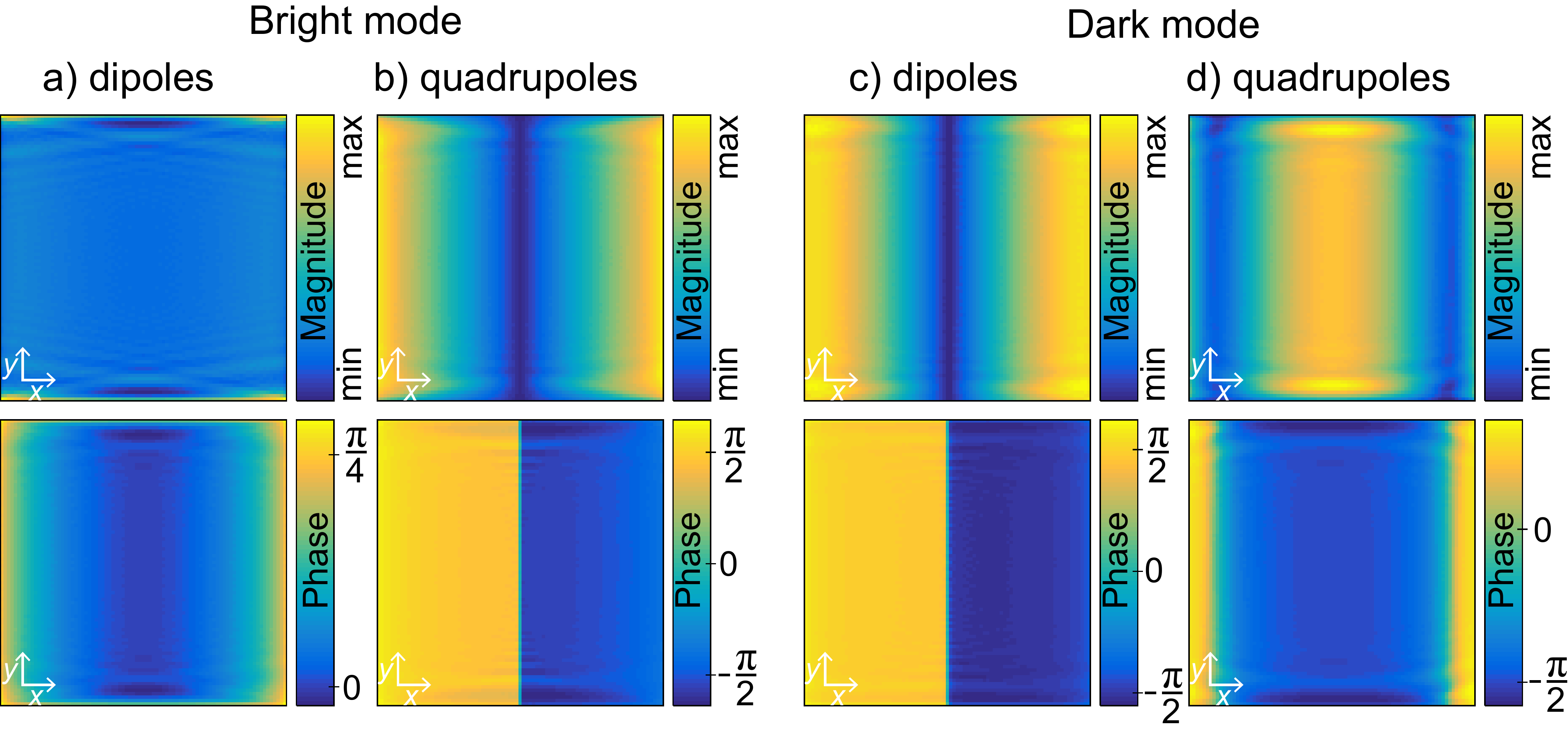}
\caption{{\it Electric dipole and quadrupole response of a $85\times85$ nanoparticle lattice calculated with the multiple scattering T-matrix method.} The nanoparticles are excited with a $y$-polarized plane wave coming from the normal
direction (a, b), or a combination of two $y$-polarized plane wave
coming from the opposite angles so that the exciting field has a $\pi$-phase
difference between the vertical edges (c, d). The upper row shows the polarization
magnitude, the lower row shows its phase. The spacing between particles
is $375\,\mathrm{nm}$ and the frequencies correspond to the bright (a, b) and dark (c, d) modes. \label{fig:multipole response dark}}
\end{figure}

\begin{figure}
\includegraphics[width=0.8\columnwidth]{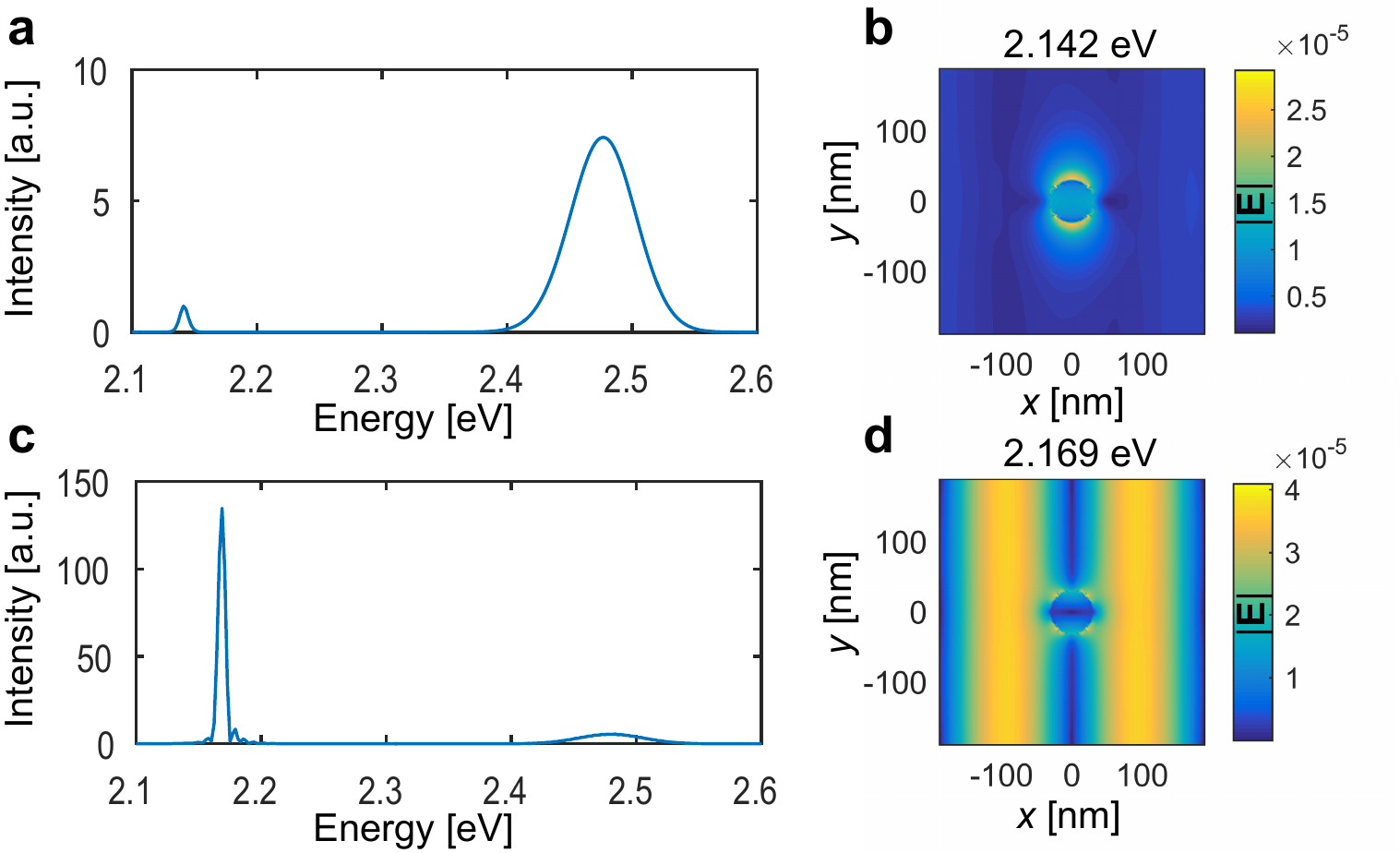}

\caption{{\it Normalized lasing spectra and corresponding field profiles obtained from FDTD simulations with gain.} The results for the bright SLR mode are shown in (a-b) and for the dark SLR mode in (c-d). The colorscales represent the electric field profiles at the lasing frequency in a single unit cell. The pump power was set to $0.63\, {\rm mJ \cdot cm^{-2}}$ in the simulation. For this pump power, both modes are well above the lasing threshold, see Fig.~2~g of the main text, but the dark mode output power is about an order of magnitude higher than that of the bright mode. In (a), the spectrum is calculated from the electric fields at the bottom left corner of the unit cell. At this location, the dark mode field profile (d) has a node, and thus it is not visible in the calculated spectra. The smaller peak at $2.142\, {\rm eV}$ corresponds to the bright SLR mode, for which the field profile is shown in (b). The larger peak at $2.48\, {\rm eV}$ comes from the pump pulse. The dark SLR mode lasing spectra is shown in (c), and the electric field profile at the lasing peak energy $2.169\, {\rm eV}$ is shown in (d). The spectra is calculated $1/4$th of the unit cell away from the corner along the horizontal axis. At this location, the bright mode has a node in its field profile, and thus is not visible in the calculated spectra. The bright SLR mode peak has been normalized to be 1 in the intensity scales. \label{fig:FDTD spectra}}
\end{figure}

\begin{figure}
\includegraphics[width=1\columnwidth]{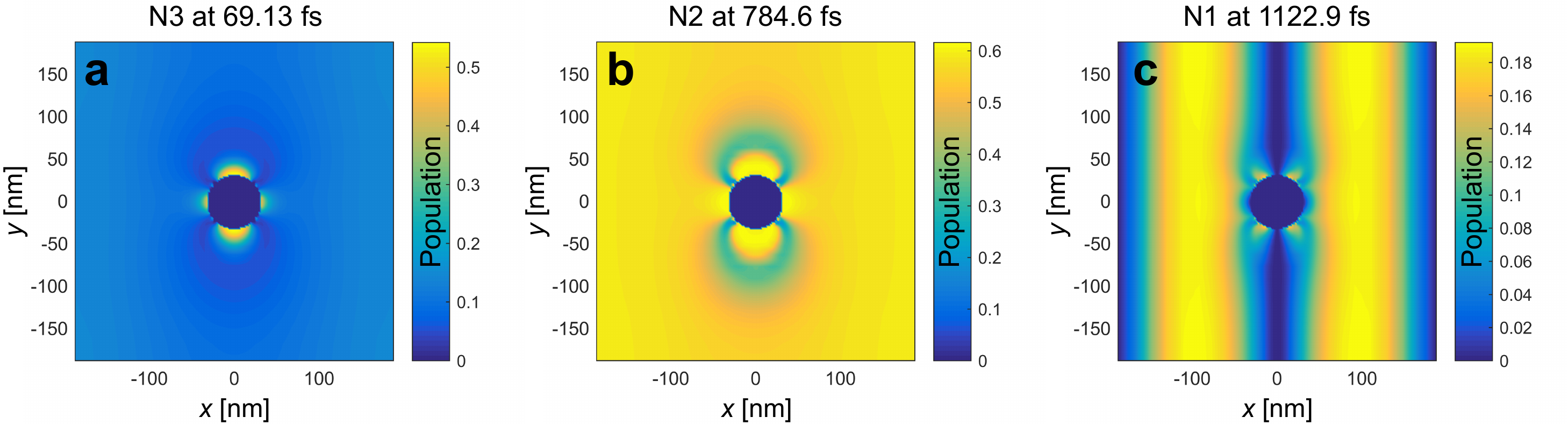}

\caption{{\it Snapshots of the populations at different points of the simulation in different levels of the 4-level gain medium.} The colorscale repesents the fraction of molecules in a specific state at a specific time of the simulation run. In (a) the highest energy state population is shown. The pump pulse is starting to interact with the structure, showing how the pump interacts with the gain medium both directly, and through the surface plasmon resonance of the nanoparticles. In (b), the upper lasing state population is shown, after the population pumped to the highest energy state has decayed into the upper lasing state, but before lasing in the SLR mode starts. In (c) is the lower lasing state population after the onset of lasing in the SLR mode. The lower lasing state population shows where stimulated emission is taking place, and this matches well with the field profile of the dark SLR mode shown in Supplementary Fig.~\ref{fig:FDTD spectra}~(d). The pump power is chosen to be below the threshold for the bright SLR mode lasing (field profile shown in Supplementary Fig.~\ref{fig:FDTD spectra} (b)). \label{fig:FDTD populations}}
\end{figure}

\begin{figure}[hb]
\includegraphics[width=0.55\columnwidth]{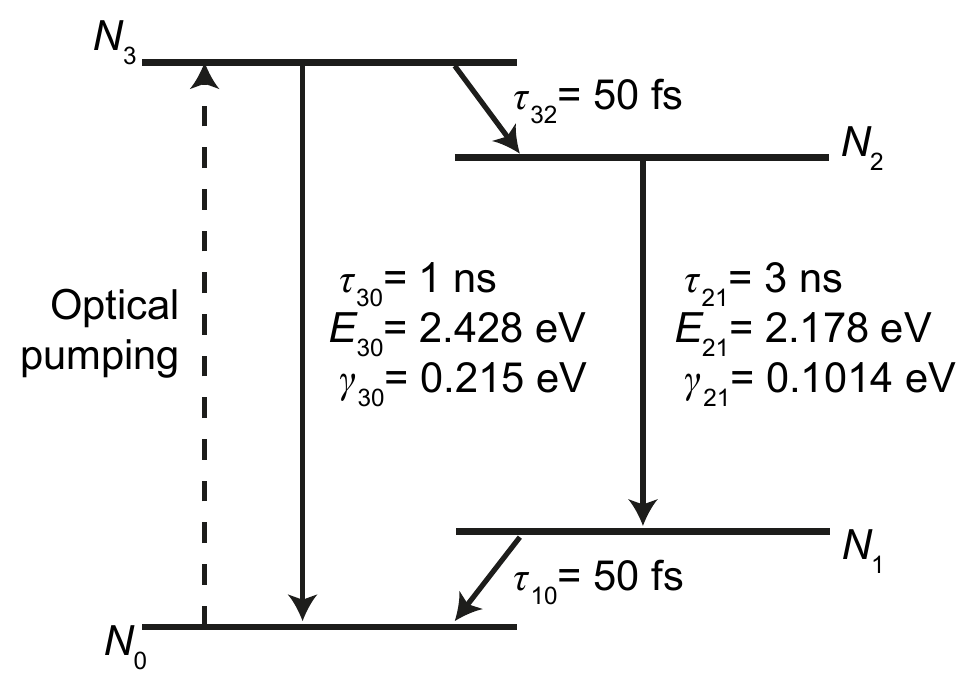}

\caption{{\it Energy level diagram and the parameters used for modelling the dye molecules in FDTD simulations.} The transition energies $E_{30}$ and $E_{21}$, and the full-widths at half-maximum of these transitions $\gamma_{30}$ and $\gamma_{21}$ are obtained from absorption and emission measurements of the Rhodamine 6G solution used in the experiments. In the beginning of a simulation, $N_0 = 1$. Then a fraction of the molecules gets excited by a pump pulse, causing level 3 to gain population. As the transition from level 3 to level 2 is much faster than the transition form level 3 to 0, the excitations quickly decay to level 2. This causes a population inversion for the transition from level 2 to 1, which supplies the necessary gain for lasing in the SLR modes. \label{fig:FDTD levels}}
\end{figure}

\begin{figure}[hb] 
\includegraphics[width=0.6\columnwidth]{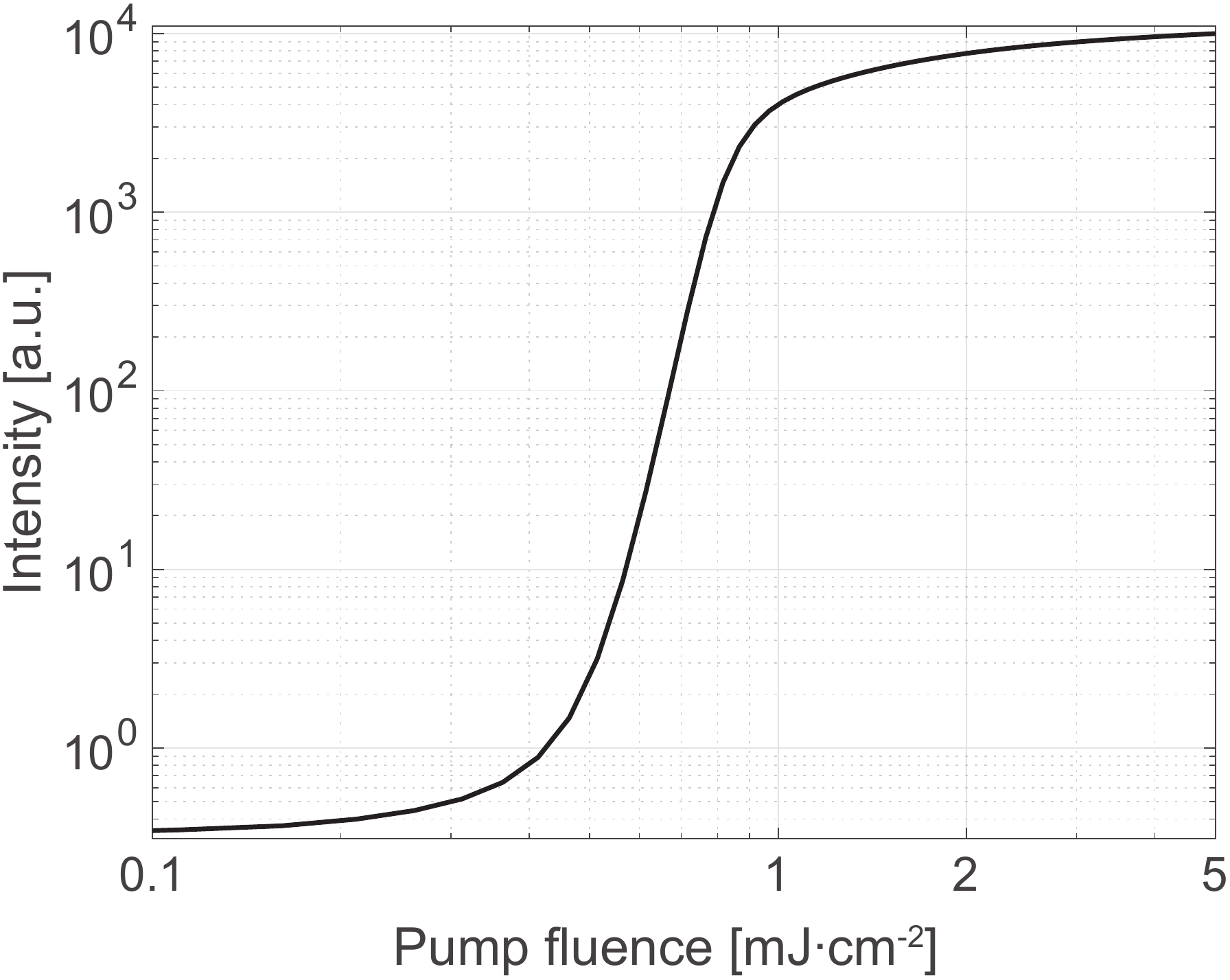}
\caption[FigSNeoC]{{\it The threshold behaviour as predicted by the neo-classical lasing model.}}
\label{fig:sup_neoclassical}
\end{figure} 

\clearpage

\subsection{Supplementary Note~1. \enspace Control experiments for the observed lasing action}

The crucial role of the plasmonic near fields for the lasing action is also demonstrated by our control experiment where the gain medium is spatially separated from the array by a poly vinyl alcohol (PVA) layer: A sample with a 100 nm PVA spacer layer still exhibited lasing action with a clear threshold, but a sample with a 500 nm PVA layer did not, see Supplementary Fig.~\ref{fig:PVA_spacer_thresholds}. Supplementary Fig.~\ref{fig:PVA_spacer_thresholds} (b) displays clear lasing in the 100 nm spacer layer sample, while the 500 nm spacer layer sample in Supplementary Fig.~\ref{fig:PVA_spacer_thresholds} (c) shows emission in a broad range of energy and momentum values.  After removing the PVA layer and consequent replacement of the gain medium, however, the sample recovered its lasing action.

The lasing action in our samples is different from distributed feedback
lasers consisting of a waveguide, gain medium and a periodic metallic
structure. In such systems, the periodic structure merely provides
the feedback for the waveguide mode. In our case, however, the refractive
index of the gain medium is closely matched with the substrate and
the cover slip, so the structure is not expected to support any waveguide
modes, but rather only the surface lattice resonance (SLR). In \cite{schokker_statistics_2015},
waveguide modes were hybridized with SLR modes, and up to 95~\% of
the particles could be removed while still preserving the lasing action.
In contrast, our samples exhibit no lasing if 75~\% (or more) of the
particles are removed. In Supplementary Figs.~\ref{fig:removed particles0}--\ref{fig:removed particles75} are shown luminescence data for samples having 0~\%, 50~\% and 75~\% of particles removed, respectively. For each sample, we plot the luminescence data with pump power below, at and above the threshold value $P_{\mathrm{TH}}$ which is
the threshold for the sample with
0~\% of particles removed. A sample having 50~\% of particles removed still lases (Supplementary Fig.~\ref{fig:removed particles50}). A sample having 75~\% of particles removed did not lase at all, even with pump powers up to $3.5\, P_{\mathrm{TH}}$ (Supplementary Fig.~\ref{fig:removed particles75}).

A sample with the same number, shape and area of nanoparticles but
in random positions shows no lasing either, even when pumped with
powers up to $P=3.5\, P_{\mathrm{TH}}$, where $P_{\mathrm{TH}}$ is
the threshold for the sample with
no particles removed. This rules out random lasing as the possible
origin for the observed lasing action.

\subsection{Supplementary Note~2. \enspace Coupled dipole and multipole scattering simulations}


\global\long\def\vect#1{\mathbf{#1}}

We simulated the linear response of our nanoparticle arrays with the
multiple-scattering $T$-matrix method, principles of which we briefly
sketch in the following. The technical details can be found in the
literature \cite{mishchenko_t-matrix_1999,xu_calculation_1996,xu_radiative_2003,taylor_optical_2011}.

Scattering of a time-harmonic electromagnetic incident field on a
single scatterer in otherwise isotropic background medium can be described
in terms of the regular $\vect M_{mn}^{(1)},\mbox{\ensuremath{\vect N}}_{mn}^{(1)}$
and outgoing $\vect M_{mn}^{(3)},\mbox{\ensuremath{\vect N}}_{mn}^{(3)}$
vector spherical wave functions (VSWFs), which are the solutions of
vector Helmholtz equation. The outgoing VSWFs $\vect M_{mn}^{(3)}$ and $\mbox{\ensuremath{\vect N}}_{mn}^{(3)}$
correspond to the field emitted by harmonically oscillating magnetic
and electric multipoles, respectively. Outside a sphere surrounding the scatterer, the electromagnetic field is the sum of the incident and scattered
field
\begin{eqnarray}
\vect E & = & \vect E_{\mathrm{inc}}+\vect E_{\mathrm{scat}}\\
\vect E_{\mathrm{inc}}(\vect r) & = & \sum_{n=1}^{\infty}\sum_{m=-n}^{n}\left[q_{mn}\vect M_{mn}^{(1)}(kr,\theta,\phi)+p_{mn}\mbox{\ensuremath{\vect N}}_{mn}^{(1)}(kr,\theta,\phi)\right],\\
\vect E_{\mathrm{scat}}\left(\vect r\right) & = & \sum_{n=1}^{\infty}\sum_{m=-n}^{n}\left[b_{mn}\vect M_{mn}^{(3)}(kr,\theta,\phi)+a_{mn}\mbox{\ensuremath{\vect N}}_{mn}^{(3)}(kr,\theta,\phi)\right],
\end{eqnarray}
where $k$ is the wavenumber of the field in the background medium
at a given frequency. Because of the linearity of Maxwell's equations,
there is a linear relation between the expansion coefficients of the
incident and scattered parts
\begin{eqnarray}
a_{mn} & = & \sum_{n'=1}^{\infty}\sum_{m'=-n'}^{n'}\left[T_{mnm'n'}^{EE}p_{m'n'}+T_{mnm'n'}^{EM}q_{m'n'}\right],\\
b_{mn} & = & \sum_{n'=1}^{\infty}\sum_{m'=-n'}^{n'}\left[T_{mnm'n'}^{ME}p_{m'n'}+T_{mnm'n'}^{MM}q_{m'n'}\right].
\end{eqnarray}
The coefficients $T_{mnm'n'}^{tt'}$ are the elements of the T-matrix
which completely characterizes the linear electromagnetic response
of a single scatterer.

Using the translation addition theorem, the outgoing VSWFs can be
re-expanded into the regular VSWFs in a spherical coordinate system
with a different origin (labeled with $j$, as opposed to the original
system which we label with $l$) 
\begin{eqnarray}
\vect M_{\mu\nu}^{(3)}(\vect r_{l}) & = & \sum_{nm}\left[A_{mn\mu\nu}^{(1)}(\vect r_{lj})\vect M_{mn}^{(1)}(\vect r_{j})+B_{mn\mu\nu}(\vect r_{lj})\vect N_{mn}^{(1)}(\vect r_{j})\right],\\
\vect N_{\mu\nu}^{(3)}(\vect r_{l}) & = & \sum_{mn}\left[B_{mn\mu\nu}^{(1)}(\vect r_{lj})\vect M_{mn}^{(1)}(\vect r_{j})+A_{mn\mu\nu}(\vect r_{lj})\vect N_{mn}^{(1)}(\vect r_{j})\right],\\
 &  & \left|\vect r_{j}\right|<\left|\vect r_{lj}\right|,
\end{eqnarray}
where the vector $\vect r_{lj}=\vect r_{l}-\vect r_{j}$ connects
the origins of both systems. If a second scatterer is present in the
centre of the new coordinate system, the field scattered
from the first scatterer incident onto the second one can be obtained.
This way one can proceed with an arbitrary number of scatterers, leading
to a linear system that, in matrix notation, is given by
\begin{equation}
\begin{bmatrix}\vect a^{j}\\
\vect b^{j}
\end{bmatrix}=\mbox{\ensuremath{\vect T}}^{j}\left(\begin{bmatrix}\vect p^{j0}\\
\vect q^{j0}
\end{bmatrix}+\sum_{l\ne j}\begin{bmatrix}\vect A(k\vect r_{lj}) & \vect B(k\vect r_{lj})\\
\vect B(k\vect r_{lj}) & \vect A(k\vect r_{lj})
\end{bmatrix}\begin{bmatrix}\vect a^{l}\\
\vect b^{l}
\end{bmatrix}\right),
\end{equation}
which can be readily solved to obtain the solution for the scattering
problem on a nanoparticle array.

Expressions for the translation coefficients $A_{mn\mu\nu}(\vect r_{lj}),B_{mn\mu\nu}(\vect r_{lj})$
as well as the expansion coefficients $p_{mn},q_{mn}$ for the initial
exciting plane wave can be found in the literature \cite{xu_radiative_2003,xu_calculation_1996}.
The only missing part is then finding the $T$-matrix for a single
nanoparticle, which we compute by the boundary element method \cite{SCUFF1,SCUFF2}.

The simulations presented in the article were performed up to the
quadrupole order, i.e. only the elements with $n=1,2$ were included
in the calculation. Including higher-order terms while keeping the
array size reasonably large is problematic mainly due to the memory
limitations.


In Supplementary Fig.~\ref{fig:multipole response dark} we show the simulated electromagnetic
response of a $85\times85$ particle lattice at both bright and dark mode energies. We note that experimentally we observed the bright mode lasing perpendicularly to the array plane, while the dark mode had two beams coming out with slight deviation from the sample normal (see also Fig.~2~c of the manuscript). Therefore, in order to excite these modes with an external plane wave, one has to use direct incidence for the bright mode, and two plane waves with a slightly tilted angle and a $\pi$ phase shift for the dark mode. 

For the bright mode, the lattice is excited by
a $y$-polarized plane wave coming from the normal direction: the electric dipole moments obtain approximately constant magnitude and phase across the array (Supplementary Fig.~\ref{fig:multipole response dark} (a)). At the center of the array, the standing wave due to two counter propagating radiation fields creates a standing wave pattern which has an antinode at each particle location. Due to symmetry, such a standing wave cannot excite the quadrupole moment into the center particles, resulting to a minimum magnitude of the quadrupolar moment at the array center (Supplementary Fig.~\ref{fig:multipole response dark} (b)). At the edges, however, only one of the counter propagating is present. Such a wave can excite the quadrupole moment to finite size of the particles (diameter 60 nm). The left and right propagating plane waves are out of phase by $\pi$, creating an abrupt phase shift at the center of the array (Supplementary Fig.~\ref{fig:multipole response dark}~(b) bottom). 

We then inspect the response at the dark mode energy (Supplementary Figs.~\ref{fig:multipole response dark} (c, d)), where we use an incidence angle of $\arcsin(\pi/k d_x)= 5.87 \times 10^{-3}$ rad. Here $d_x$ is the array size. At the center of the array, the two counter propagating plane waves are of equal magnitude and out of phase by $\pi$, creating a standing wave node at each particle location. Due to symmetry, such a wave cannot excite the dipolar moment into the particles, resulting to 1) minimum of the dipole moment and 2) maximum of the quadrupole moment at the center of the array. At the edges, only one of the counter propagating plane waves is present, therefore exciting the dipole moment of the particle. As the counter propagating plane waves are out of phase by $\pi$, the dipole moment distribution undergoes an abrupt phase shift at the center of the array. 

One can summarize the results in the following way: 1) For the bright mode, the dipolar moment maximizes at the center and has roughly a constant phase across the array. 2) Quadrupolar moments minimize, and undergo a $\pi$ phase shift at the center of the array. 3) For the dark mode, the opposite happens, i.e., dipoles minimize and undergo a $\pi$ phase shift at the center of the array and 4) quadrupoles maximize at the center and have roughly constant phase across the array.

\subsection{Supplementary Note~3. \enspace FDTD simulations with an active gain medium}

The response of the dye molecules placed on top of the samples in the experiments was simulated using the FDTD method, similar to Refs. \cite{DridiSchatz2013paper} and \cite{dridi_lasing_2015}. We use Lumerical's FDTD Solutions software, version 8.16.903 for the simulations. The dye molecules are modelled as 4-level systems, which are coupled in time domain to the electric fields of the nanoparticle arrays. The dye is placed as a 150nm thick film on top of the particles. The 4-level energy diagram, and parameters used for the simulations are shown in Supplementary Fig.~\ref{fig:FDTD levels}. The dye transition energies and full-width half-maximum values are obtained from absorption and emission measurements of the 31 mM Rhodamine 6G dye solution used in the experiments. The density of the emitters is $1.867\cdot 10^{19} \, {\rm cm^{-3}}$, corresponding to the $31\, {\rm mM}$ concentration used in the experiments.

In the beginning of the simulation, all of the dye molecules are in the ground state, $N_0=1$. We then use a $30\, {\rm fs}$ pump pulse, centered at $500\, {\rm nm}$ ($2.48\, {\rm eV}$), to excite the molecules. The pump is linearly polarized along the $y$-axis. Initially this pulse excites a fraction of the molecules to state $N_3$, which then relax to state $N_2$, creating a population inversion. After a sufficient population inversion is reached, stimulated emission to the dark or bright SLR modes quickly brings down the population of the state $N_2$. Emission spectrum obtained for the dark and bright modes from the FDTD simulations are shown in Supplementary Fig.~\ref{fig:FDTD spectra}, together with the field profiles at peak emission energies from the same simulation run. The energies of these modes are slightly shifted compared to the simulations without the active media (see Fig.~1 of the main text) through a change in the refractive index, which depends on the input power. The lattice period is $375\, {\rm nm}$ for both $x$ and $y$ directions.

Snapshots of the level populations observed with a pump power of $0.31\, {\rm mJ \cdot cm^{-2}}$ at different times are shown in Supplementary Fig.~\ref{fig:FDTD populations}. The level populations are monitored in the $x$-$y$ plane of the simulation, at the middle height of the nanoparticles. In Supplementary Fig.~\ref{fig:FDTD populations}~(a) is shown the population of state $N_3$ slightly before the pump pulse reaches its maximum intensity. Due to the localized surface plasmon resonance in the nanoparticles, the dye molecules close to the particle are pumped the hardest. In Supplementary Fig.~\ref{fig:FDTD populations}~(b) the upper lasing level $N_2$ population is shown after the pump has passed through the structure, and the population in level $N_3$ has decayed into level $N_2$. The structure does not lase yet. Slightly away from the nanoparticle, a band with lower amount of excited molecules is observed. This is due to overpumping the dye molecules close to the particle through the surface plasmon resonance. In Supplementary Fig.~\ref{fig:FDTD populations}~(c) is shown the lower lasing state population $N_1$ after the structure has started to lase in the dark SLR mode. The lower lasing state population shows where stimulated emission has occurred, and this can be compared to the field profile of the dark SLR mode shown in Supplementary Fig.~\ref{fig:FDTD spectra}~(d).

\subsection{Supplementary Note~4. \enspace Neo-classical model of lasing}
We model lasing in our system also along the neo-classical lines as explained in Ref.~\cite{martikainen_modelling_2016}. All nano-particles are assumed to be equal and well described by the same dipole.  The array structure influences the effective polarizability via the coupled dipole approximation (CDA). Oscillating dipoles can then give 
rise to the lasing field when they are coupled with a gain medium. We model the Rhodamine 6G gain medium as a four-level system with parameters taken from Ref.~\cite{cuerda_theory_2015}. 
The transition between the ground state and highest state is driven by a coherent laser pulse similar to experiments. We assume $10\,\%$ active molecules, a Purcell factor of $2$, and SLR loss rate of $\gamma_{SLR}=3\,{\rm meV}$. In Supplementary Fig.~\ref{fig:sup_neoclassical} we demonstrate the resulting threshold behaviour. These results show that a neo-classical model that describes the SLR modes in a simplified manner is sufficient to produce the overall threshold behaviour. However, it is not able to describe lasing in the dark and bright modes included in the full FDTD-based model in Supplementary Note~3.

\end{document}